\documentclass[nofootinbib,aps,prd,reprint,superscriptaddress,showkeys]{revtex4-2}

\usepackage[utf8]{inputenc} 
\usepackage{graphicx,overpic,mathtools}
\usepackage{amsthm,amsmath,amssymb,hyperref}
\usepackage{braket,bm,bbm,setspace}
\usepackage[normalem]{ulem} 
\usepackage{physics}
\usepackage{float}
\usepackage[makeroom]{cancel}
\usepackage[english]{babel}
\usepackage{xcolor}
\usepackage{tensor}
\graphicspath{ {images/} }
\addto\captionsspanish{}
\hypersetup{
	colorlinks=true,
	pdfborder={0 0 0},
	citecolor=purple,
	linkcolor=blue,
	filecolor=blue,
	urlcolor=blue,
}
\usepackage{subfigure}

\usepackage{orcidlink}

\usepackage{tikz}
\usetikzlibrary{quantikz2}

\newtheorem{lemma}{Lemma}
\newtheorem{definition}{Definition}

\makeatletter
\def\th@plain{
	\thm@notefont{}
	\itshape
}
\def\th@definition{
	\thm@notefont{}
	\normalfont
}
\makeatother

\begin{document}

\newcommand{\coin}{\widetilde{N}}
\newcommand{\twnan}{graph-phased }
\newcommand{\Twnan}{Graph-phased }
\newcommand{\vertexphased}{vertex-phased }
\newcommand{\Vertexphased}{Vertex-phased }
\newcommand{\linkphased}{link-phased }
\newcommand{\Linkphased}{Link-phased }
\newcommand{\ci}{\text{i}}
\newcommand{\adder}{\mathcal{P}^+}
\newcommand{\phimatrix}{\varphi}
\newcommand{\cs}{0.2cm}
\newcommand{\css}{0.3cm}
\newcommand{\rs}{0.3cm}
\newcommand{\pseudoprojector}{\Sigma}
\newcommand{\link}{link }
\newcommand{\Link}{Link }
\newcommand{\update}{V}
\newcommand{\apriori}{\textit{a priori} }
\newcommand{\Apriori}{\textit{A priori} }

\title{Complex-Phase Extensions of Szegedy Quantum Walk on Graphs}
\author{Sergio A. Ortega \orcidlink{0000-0002-8237-7711}}
\email{sergioan@ucm.es}
\affiliation{Departamento de Física Teórica, Universidad Complutense de Madrid, 28040 Madrid, Spain}
\author{Miguel A. Martin-Delgado \orcidlink{0000-0003-2746-5062}}
\email{mardel@ucm.es}
\affiliation{Departamento de Física Teórica, Universidad Complutense de Madrid, 28040 Madrid, Spain}
\affiliation{CCS-Center for Computational Simulation, Universidad Politécnica de Madrid, 28660 Boadilla del Monte, Madrid, Spain.}

\begin{abstract}
	{This work introduces a \twnan Szegedy's quantum walk, which incorporates \link phases and local arbitrary phase rotations (APR), unlocking new possibilities for quantum algorithm efficiency. We demonstrate how to adapt quantum circuits to these advancements, allowing phase patterns that ensure computational practicality. The \twnan model broadens the known equivalence between coined quantum walks and Szegedy's model, accommodating a wider array of coin operators. Through illustrative examples, we reveal intriguing disparities between classical and quantum interpretations of walk dynamics. Remarkably, local APR phases emerge as powerful tools for marking graph nodes, optimizing quantum searches without altering graph structure. We further explore the surprising nuances between single and double operator approaches, highlighting a greater range of compatible coins with the latter. To facilitate these advancements, we present an improved classical simulation algorithm, which operates with superior efficiency. This study not only refines quantum walk methodologies but also paves the way for future explorations, including potential applications in quantum search and PageRank algorithms. Our findings illuminate the path towards more versatile and powerful quantum computing paradigms.}
\end{abstract}


\maketitle

\section{Introduction}\label{Introduction}

Quantum walks are algorithms born from the quantization of classical Markov chains. They were first proposed in the discrete time version \cite{QRW}, and later in continuous time \cite{Trees}. The same as classical walks, quantum walks have given rise to a wide variety of algorithms for problems such us triangle finding \cite{Triangles}, element distinctness \cite{ED} and quantum search \cite{QRW_Search}.

Quantum walks in discrete time usually required an inner degree of freedom, giving rise to what is called the coined quantum walk model \cite{Coined-general-graph}. These walks act on undirected graph, and have difficulties quantizing arbitrary Markov chains. To solve this issue, Szegedy introduced a coinless quantum walk on bipartite graphs quantizing a general Markov chain on weighted graphs by a duplication process \cite{Szegedy}. This quantum walk has been shown to be useful for problems of optimization \cite{Lemieux,Qfold,QMS,GWQMA,qBIRD}, testing graph completeness \cite{Giordano}, classification \cite{Paparo1,Paparo2,APR,Randomized}, quantum search \cite{Portugal,Searchrank,S_queries}, and machine learning \cite{Paparo3}.

Soon after Szegedy's work, an alternative version of Szegedy's quantum walk avoiding the need to duplicate the graph \cite{Notes} was developed. This formulation produces more general results for classical weighted graphs \cite{Semiclassical}, and moreover, allows establishing an equivalence between Szegedy's quantum walk and the coined model \cite{Sandbichler-master,Wong_1,Wong_2}.

Different modifications of Szegedy's quantum walk introducing complex phases have been proposed in the context of Grover walks \cite{Twisted}, staggered quantum walks \cite{Staggered} and the quantum PageRank algorithm \cite{APR}, where they have provided algorithms outperforming the standard walk. In this work we provide a further generalization of the phase extensions, obtaining what we call the \twnan Szegedy's quantum walk. Moreover, we have studied how provided an efficient quantum circuit for the implementation of Szegedy's quantum walk \cite{Q_circuits,Q_circuits_2}, we can modify it to include the complex phase extensions.

An important result we find for our \twnan Szegedy's model is that the set of coined quantum walks that can be cast into a Szegedy's walk is increased considerably. We review the conditions that must satisfy the coin operators \cite{Sandbichler-master}, and provide new conditions considering also the phase extensions. Furthermore, we also show how the new extensions that we introduce in our work allow marking nodes in a graph using oracles. We also use this extension to review the proof that the method of absorbing vertices usually used for marking \cite{Szegedy,Portugal} is equivalent to the use of a different coin in the coined walk \cite{QRW_Search,Wong_1}.

Finally, we provide an efficient classical simulation algorithm for the \twnan Szegedy's quantum walk considering all possible phase extensions. For this, we have modified a recent algorithm that we proposed for the current Szegedy's walks, which scales as $\mathcal{O}(N^2)$ in both time and memory requirements \cite{Squwals}. This algorithm outperforms previous simulation methods that require the explicit construction of the unitary evolution matrix, scaling as $\mathcal{O}(N^3)$ for dense weighted graphs.

This paper is structured as follows. In Section \ref{sec:Szegedy} we review the formulation of Szegedy's quantum walk to later introduce the \twnan model. In Section \ref{sec:Circuits} we show the efficient construction of quantum circuits. In Section \ref{sec:Coin} we study the equivalence between the \twnan Szegedy's walk and the coined model. In section \ref{sec:Line} we show examples of coined quantum walks on the infinite line. In section \ref{sec:Marking} we show how some phase extensions can be used for marking nodes in a graph. In section \ref{sec:SQUWALS} we present the classical simulation algorithm for the \twnan walk considering all the phase extensions. Finally, we summarize and conclude in Section \ref{Conclusions}.

\section{\Twnan Szegedy Quantum Walk}\label{sec:Szegedy}

In this section we review the standard quantization of Szegedy's quantum walk \cite{Szegedy,Notes} and the current extensions found in the literature \cite{Twisted,Staggered,APR}, to later generalize it introducing the \twnan Szegedy's model.

\subsection{Szegedy's quantum walk}

A classical Markov chain is a stochastic process that can be represented as a random walk on the nodes of a graph. For a weighted graph with $N$ nodes, this walk is described by a $N\times N$ transition matrix $G$. By convention in this work we choose that the transition matrix is column-stochastic, so that all the columns add up to $1$. Thus, the elements $G_{ji}$ are the probabilities of the walker jumping from node $i$ to node $j$. Let us define $p(t)$ as a $N$-dimensional column vector whose elements are the probabilities of the classical walker being at each node at time step $t$. Then, the classical walk evolution is governed by the following equation:
\begin{equation}\label{classical_evolution}
	p(t) = G^t p(0).
\end{equation}

The classical Markov chain is quantized through Szegedy's quantum walk \cite{Szegedy,Notes}. In this quantum walk the Hilbert space is the span of all the vectors representing the $N \times N = N^2$ directed edges of the graph, i.e.,
\begin{equation}\label{hilbert_szegedy}
	\mathcal{H_S} = \text{span}\lbrace\left|i\right>_1\left|j\right>_2,\ i,j = 0,1,...,N-1\rbrace,
\end{equation}
where the states with indexes $1$ and $2$ refer to the nodes on two copies of the original graph. Thus, the states are defined over two quantum registers. In this paper we count the nodes of the network, and therefore the matrix indexes, from $0$ to $N-1$. We define the vectors
\begin{equation}\label{psi_i}
	\left|\psi_i\right> := \left|i\right>_1 \otimes \sum_{k=0}^{N-1} \sqrt{G_{ki}}\left|k\right>_2,
\end{equation}
which are a superposition of the vectors representing the edges outgoing from the $i$-th vertex, whose coefficients are given by the square root of the $i$-th column of the matrix $G$. From these vectors we define the following projector operator:
\begin{equation}\label{projector}
	\Pi := \sum_{i=0}^{N-1} \left|\psi_i\right>\left<\psi_i\right|,
\end{equation}
and use it to define a reflection operator on the subspace generated by the $\left|\psi_i\right>$ states:
\begin{equation}\label{reflection}
	R := 2\Pi - \mathbbm{1}.
\end{equation}
The quantum walk evolution operator $U_s$ is defined as
\begin{equation}\label{U}
	U_s := S_wR,
\end{equation}
where $S_w$ is a swap operator that exchanges the two quantum registers, i.e.,
\begin{equation}\label{swap}
	S_w := \sum_{i,j=0}^{N-1} \left|i,j\right>\left<j,i\right|.
\end{equation}

The Szegedy's quantum walk described in this manner is actually an alternative definition \cite{Notes} that generalizes the original one \cite{Szegedy} and is more suitable for establishing an equivalence with the coined quantum walk model \cite{Coined-general-graph}, which we will explore further in section \ref{sec:Coin}. However, the original formulation made by Szegedy consisted on a coinless quantum walk based on two reflections, so that the quantum walk evolution operator was defined as $W_s = R_BR_A$. It turns out that $R_A = R$ and $R_B = S_wRS_w$, so that the original unitary operator $W_s$ corresponds to two steps of the alternative version $U_s$, i.e., $W_s = U_s^2$. Thus, we will refer to it as the double Szegedy operator.

Finally, the initial state is usually constructed by a superposition of the $\left|\psi_i\right>$ states, and the probability distribution of the walker after each time step of the quantum walk is usually obtained measuring the first register. However, there are also algorithms where the information of interest is obtained measuring the second register instead \cite{Paparo1,Paparo2,APR,Searchrank,Randomized}.

\subsection{Complex phase extensions}

An extension of Szegedy's model, dubbed twisted Szegedy walk, introduced complex phases by means of a kind of complex-valued weights \cite{Twisted}. In our notation, this corresponds to an extended Szegedy walk where the $\left|\psi_i\right>$ states are substituted by \cite{Staggered}
\begin{equation}\label{psi_i_extended}
	\left|\psi_i(\varphi)\right> := \sum_{k=0}^{N-1} e^{\ci\varphi_{ik}} \sqrt{G_{ki}}\left|i\right>_1\left|k\right>_2,
\end{equation}
where $\varphi$ is an $N\times N$ matrix, so that the element $\varphi_{ij}$ is the phase associated to the edge state $\left|i\right>_1\left|j\right>_2$. Thus, the unitary evolution operator is $U_s(\varphi) = S_wR(\varphi)$, where $R(\varphi)$ reflects around the subspace generated by the $\left|\psi_i(\varphi)\right>$ states. When all the phases are $0$ the standard Szegedy's walk is recovered. Note that due to the convention of the transition matrix $G$ being column-stochastic, the weight probability associated to edge state $\left|i\right>_1\left|j\right>_2$ is $G_{ji}$. Thus, the real weights in the transition matrix $G$ are associated with the elements of $\varphi^T$, and vice versa.

Another modification of Szegedy's quantum walk changing the reflection $R$ by an arbitrary phase rotation (APR) was proposed in the context of quantum PageRank \cite{APR}. Whereas in the previous case $N^2$ phases were introduced, in this case a single phase denoted as $\theta$ is used to define the phase rotation operator
\begin{equation}\label{single-apr}
	R(\theta) := (1-e^{\ci\theta})\Pi - \mathbbm{1},
\end{equation}
so that the unitary evolution operator is $U_s(\theta) = S_wR(\theta)$. The standard walk is recovered for $\theta = \pi$.

In order to differentiate both phase extensions, we will call \link phases to the phases $\varphi_{ij}$ associated to the edges of the graph, and APR phase to the phase $\theta$ related to the arbitrary phase rotation. These two kind of extensions are compatible and can be applied at the same time. Before constructing a general operator that combines all the phases, it is important to mention that the twisted Szegedy walk provided another extension with a complex phase for each edge acting inside the swap operator \cite{Twisted}. Nevertheless, this is beyond the scope of this work, as we are interested in the extensions of the reflection operator $R$, which acts as the coin of the coined quantum walk. Moreover, we can still extend further this operator.

\subsection{\Twnan Szegedy's quantum walk}

The APR phase $\theta$ acts in a global manner in the graph. However, the same as there are $N^2$ phases $\varphi_{ij}$ associated to the edges of the graph, we could think of $N$ APR phases $\theta_i$ associated to each of the nodes of the graph. Let us expand the projector $\Pi$ in \eqref{single-apr}, and introduce the constant factor $(1-e^{\ci\theta})$ in the sum:
\begin{equation}
	R(\theta) = \sum_{i=0}^{N-1} (1-e^{\ci\theta})\left|\psi_i\right>\left<\psi_i\right| - \mathbbm{1}.
\end{equation}
Now, we can give $\theta$ a different value $\theta_i$ for each node $i$. Let us define $\vec{\theta}$ as a vector with the $N$ different APR phases. The phase rotation operator becomes
\begin{equation}\label{multi-apr}
	R(\vec{\theta}) = \sum_{i=0}^{N-1} (1-e^{\ci\theta_i})\left|\psi_i\right>\left<\psi_i\right| - \mathbbm{1}.
\end{equation}
These APR phases act locally in the nodes of the graph, and the sum cannot be factorized with a projector operator.

Now we can define the \twnan model joining all the phase extensions. Let us define the following operator:
\begin{equation}\label{sigma}
	\pseudoprojector(\vec{\theta},\varphi) := \frac{1}{2}\sum_{i=0}^{N-1} (1-e^{\ci\theta_i})\left|\psi_i(\varphi)\right>\left<\psi_i(\varphi)\right|.
\end{equation}
This operator generalizes the projector $\Pi$ in \eqref{projector}, which is recovered for $\theta_i = \pi$ and $\varphi_{ij} = 0$. It is a kind of pseudoprojector, since acting on an arbitrary state provides a state in the subspace spanned by the $\left|\psi_i(\varphi)\right>$ states. However, it does not corresponds to the actual orthogonal projection.

The phase rotation associated to this walk is defined as
\begin{equation}\label{general_R}
R(\vec{\theta},\varphi) := 2\pseudoprojector(\vec{\theta},\varphi) - \mathbbm{1}.
\end{equation}
The unitary evolution operator is defined as
\begin{equation}
	U_s(\vec{\theta},\varphi) := S_wR(\vec{\theta},\varphi).
\end{equation}

Again, we can think of the double Szegedy operator $W_s$ using two times the operator $U_s(\vec{\theta},\varphi)$. In general the phases of both operators can be different, so that the most general double Szegedy operator is:
\begin{equation}
	W_s(\vec{\theta}_1,\phimatrix_1,\vec{\theta}_2,\phimatrix_2) = U_s(\vec{\theta}_2,\phimatrix_2)U_s(\vec{\theta}_1,\phimatrix_1).
\end{equation}

Before going on, let us fix some definitions about the different models we will deal with in this work.

\begin{definition}[Standard Szegedy's quantum walk]\label{def:szegedy-standard}
It is the Szegedy's quantum walk without any phase extension.
\end{definition}

\begin{definition}[\Linkphased Szegedy's quantum walk]\label{def:szegedy-link}
It is the Szegedy's quantum walk with \link phases $\varphi_{ij}$, but without APR phases.
\end{definition}

\begin{definition}[Szegedy's quantum walk with global APR]\label{def:szegedy-global-apr}
It is the Szegedy's quantum walk with the same APR phase $\theta$ for all nodes, but without \link phases.
\end{definition}

\begin{definition}[\Vertexphased Szegedy's quantum walk]\label{def:szegedy-local-apr}
It is the Szegedy's quantum walk with multiple local APR phases $\theta_i$, but without \link phases.
\end{definition}

\begin{definition}[\Twnan Szegedy's quantum walk]\label{def:szegedy-graph}
It is the most general Szegedy's quantum walk with all the phase extensions.
\end{definition}

Unless otherwise stated, we will consider the single step operator $U_s$ when we refer to each model.

\section{Quantum circuits}\label{sec:Circuits}

Constructing a quantum circuit for Szegedy's quantum walk is not a trivial task since it depends on the particular weighted graph where the walk takes place. However, there is a general structure for decomposing the circuit. Here we review the construction of quantum circuits for the standard model, and later show how to modify them to include the phase extensions. Thus, provided a quantum circuit for the standard Szegedy's quantum walk, we can obtain the circuit of the \twnan model.

\subsection{Standard Szegedy's circuit}

Let us consider for simplicity that the number of nodes of the graph $N$ can be expressed as a power of two, such that $N=2^n$ for some integer $n$. Then, the quantum circuit needs two quantum registers of $n$ qubits each. For $N \neq 2^n$ we could augment the graph in a manner such that the original walk occurs in an invariant subspace of the augmented graph.

The evolution operator in \eqref{U} was $U_s = S_wR$. The swap $S_w$ is trivially a bunch of swap gates between the qubits of the first register and the qubits of the second register. However, the reflection $R$ is more complicated. This operator needs to be diagonalized \cite{Q_circuits,Q_circuits_2}. With this purpose, we define the update operator $\update$, which creates the $\left|\psi_i\right>$ state if the first register is in state $\left|i\right>_1$, provided that the second register is in $\left|0\right>_2$. Thus, its action is defined as
\begin{equation}
	\update\left|i\right>_1\left|0\right>_2 = \left|\psi_i\right>,
\end{equation}
and the action on the rest of the computational basis is irrelevant as long as it is unitary. The update operator $\update$ diagonalizes the reflection operator $R$ as $D = \update^\dagger R\update$, so that using \eqref{reflection} and \eqref{projector} we obtain:
\begin{eqnarray}
D &=& 2\sum_{i=0}^{N-1} \update^\dagger_i\left|\psi_i\right>\left<\psi_i\right|\update_i - \mathbbm{1}\nonumber\\
&=& 2\sum_{i=0}^{N-1} \left|i\right>_1\left<i\right| \otimes \left|0\right>_2\left<0\right| - \mathbbm{1}.
\end{eqnarray}
Factorizing the identity in both registers as $\mathbbm{1} = \mathbbm{1}_1\otimes \mathbbm{1}_2$, and using the completeness relation of the identity in the first register, we obtain
\begin{equation}\label{diagonal}
D = \mathbbm{1}_1 \otimes \left(\left|0\right>_2\left<0\right| - \mathbbm{1}_2\right).
\end{equation}
This operator corresponds to a reflection around the state $\left|0\right>_2$ on the second register. This is a diagonal operator with eigenvalue $+1$ for $\left|0\right>_2$ and $-1$ for the rest of the computational basis of the second register. In order to implement it, we actually use the operator $-D$, so that we end up applying the reflection $R$ up to an unimportant global phase of $-1$. Thus, we need to flip the phase of state $\left|0\right>_2$. This is accomplished with a controlled-$Z$ gate.

For the sake of generalizing later, let us instead use the phase gate $P(\theta)$:
\begin{equation}
	P(\theta) = 
	\left(\begin{array}{cc}
		1 & 0 \\
		0 & e^{\ci\theta}
	\end{array}\right),
\end{equation}
which for the standard phase $\theta = \pi$ recovers the $Z$ gate. When this gate is controlled, the target qubit can be whatever qubit since the phase $e^{\ci\theta}$ of the state cannot be attributed to a particular qubit. Thus, we use a representation in which all the qubits represent the state to whom we want to apply the phase $e^{\ci\theta}$, and they control the phase gate on a ghost target qubit \cite{DGrover}, as shown in Figure \ref{F:phase_gate}.

%

\newcommand{\cspc}{1.4cm}
\begin{figure}[hbpt]
	\centering
	\subfigure[]{\includegraphics[scale=1]{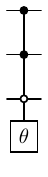}}
	\raisebox{\cspc}{=}
	\subfigure[]{\includegraphics[scale=1]{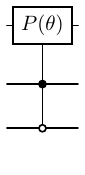}}
	\raisebox{\cspc}{=}
	\subfigure[]{\includegraphics[scale=1]{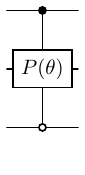}}
	\raisebox{\cspc}{=}
	\subfigure[]{\includegraphics[scale=1]{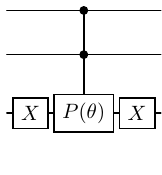}}
	\caption{a) Quantum circuit representation of a multi-controlled-$P(\theta)$ gate on a ghost qubit, which in this example applies a relative phase $e^{\ci\theta}$ to the state $\left|110\right>$. The real target can be whatever qubit as shown in b)-d), as long as the target qubit is surrounded by $X$ gates if it must be in $\left|0\right>$ for the application of the phase gate.}
	\label{F:phase_gate}
\end{figure}

\begin{figure}[hbpt]
	\centering
	\includegraphics[scale=1]{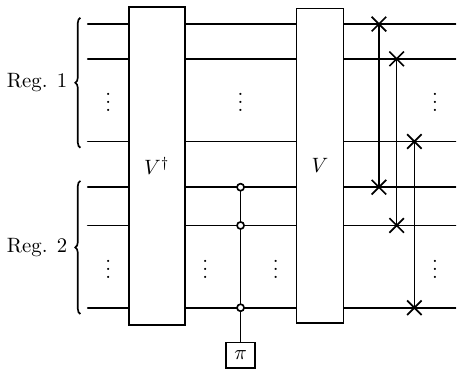}
	\caption{Quantum circuit decomposition of the standard single Szegedy unitary evolution operator $U_s=S_wR$. Each register has $n$ qubits for a graph with $N=2^n$ nodes. The reflection $R=\update D\update^\dagger$, where $D$ is a diagonal operator that implements a reflection around state $\left|0\right>_2$ in the second register.}
	\label{F:circuit-standard}
\end{figure}

For the operator $D$, all the qubits in the second register control the gate $P(\pi)$ with the condition of all qubits being in state $\left|0\right>$. Thus, the circuit for Szegedy's quantum walk operator $U_s$ can be decomposed in a general manner as shown in Figure \ref{F:circuit-standard}. For the double operator $W_s$ we would just apply twice this circuit.


The final problem is to provide a circuit for the update operator $\update$. In general, we need to codify the $N^2$ transition probabilities of matrix $G$, so that the complexity of the circuit would scale at least as $\mathcal{O}(N^2)$ for a general dense transition matrix. For sparse graphs, where there are few transition probabilities, this operator could be implemented efficiently \cite{Q_circuits_2}. Moreover, even if the matrix is dense but the graph has some symmetry properties, this operator could also be implemented efficiently. Different implementations of the update operators for graphs with symmetry can be found \cite{Q_circuits}. Supposing an efficient implementation of $\update$ is provided, in the following subsections we will show how to modify the standard circuit to include the phase extensions.

\subsection{Adding APR phases}

Let us consider the \vertexphased Szegedy's model, where the reflection $R$ is substituted by the phase rotation operator $R(\vec{\theta})$ in \eqref{multi-apr} including local APR. We diagonalize it using the update operator $\update$ as before, so that $D(\vec{\theta}) = \update^\dagger R(\vec{\theta})\update$. Then
\begin{eqnarray}
D(\vec{\theta}) &=& \sum_{i=0}^{N-1} (1-e^{\ci\theta_i}) \update^\dagger_i\left|\psi_i\right>\left<\psi_i\right|\update_i - \mathbbm{1}\nonumber\\
&=& \sum_{i=0}^{N-1} (1-e^{\ci\theta_i}) \left|i\right>_1\left<i\right| \otimes \left|0\right>_2\left<0\right| - \mathbbm{1}\nonumber\\
&=& \sum_{i=0}^{N-1} \left|i\right>_1\left<i\right| \otimes \left[(1-e^{\ci\theta_i})\left|0\right>_2\left<0\right| - \mathbbm{1}_2\right].
\end{eqnarray}
In the case of a global APR phase $\theta$, we could again factorize it as a diagonal operator acting on the second register the same as in equation \eqref{diagonal}, which would correspond to a phase rotation around the state $\left|0\right>_2$. This would have eigenvalue $-e^{\ci\theta}$ for $\left|0\right>_2$ and $-1$ for the rest. The circuit would implement $-D(\theta)$ instead, so that applies a phase $e^{\ci\theta}$ to the state $\left|0\right>_2$ letting the rest of the computational basis unchanged. Then, the circuit would be the same as the one shown in Figure \ref{F:circuit-standard} but changing the phase $\pi$ by the general phase $\theta$.


\begin{figure}[hbpt]
	\centering
	\includegraphics[scale=1]{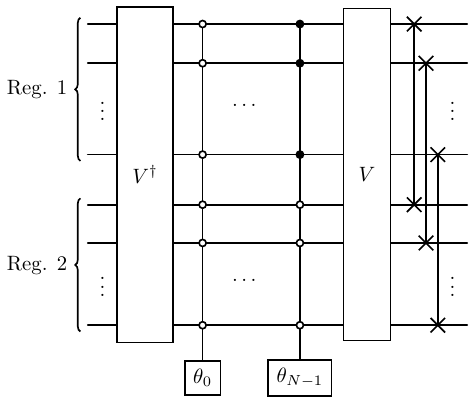}
	\caption{Quantum circuit decomposition of the single unitary operator $U_s(\vec{\theta})=S_wR(\vec{\theta})$ for the \vertexphased Szegedy's quantum. Each register has $n$ qubits for a graph with $N=2^n$ nodes. The phase rotation operator $R(\vec{\theta}) = \update D(\vec{\theta})\update^\dagger$, where $D(\vec{\theta})$ is a diagonal operator that implements $N$ different phase rotations around state $\left|0\right>_2$ in the second register. Thus, this circuit needs $N$ phase rotations controlled by the first register, from state $\left|0\right>_1$ with all qubits in $\left|0\right>$ to state $\left|N-1\right>_1$ with all qubits in $\left|1\right>$.}
	\label{F:circuit-local-apr}
\end{figure}

However, if we have local APR phases, the diagonal operator cannot be factorized. In this case the operator has the form of a uniformly controlled gate \cite{Uniformly_controlled,Uniformly_controlled_2}, which applies a different phase rotation around the state $\left|0\right>_2$ depending on the state of the first register. Thus, in general it corresponds to $N$ different phase rotations acting on the second register controlled by the $N$ states of the computational basis of the first register. The quantum circuit for this case is shown in Figure \ref{F:circuit-local-apr}.

The construction of this circuit is in general inefficient, since we need $N$ controlled phase rotation gates. However, there can be cases where a more efficient circuit is possible. The local APR phases can be distributed in some manner that the same controlled phase rotation can be used to apply it to a bunch of states at the same time. For example, if we have the same phase $\theta_e$ for even nodes, and the same phase $\theta_o$ for odd nodes, they can be implemented using only two controlled phase rotations. In this case they would be controlled by the last qubit of the first register, which indicates the parity of the state.


\begin{figure}[hbpt]
	\centering
	\includegraphics[scale=1]{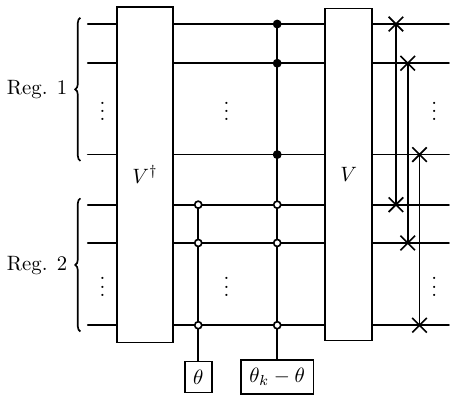}
	\caption{Quantum circuit decomposition of the single unitary operator $U_s(\vec{\theta})=S_wR(\vec{\theta})$ for the \vertexphased Szegedy's quantum walk on a graph where most nodes have the same local APR phase $\theta$, and $M$ special nodes have a different local APR phase. Each register has $n$ qubits for a graph with $N=2^n$ nodes. A phase rotation of $\theta$ is applied to all nodes, and then $M$ different phase rotations controlled by the first register are applied for the special nodes. In this example, only node $N-1$ has a different phase $\theta_k$, so that the controlled phase rotation needs all qubits in the first register to be in state $\left|1\right>$.}
	\label{F:circuit-special-apr}
\end{figure}

Another important case where an efficient implementation is possible occurs when most of the nodes have the same phase $\theta$, and there is a small set $\mathcal{M}$ of special nodes with a different phase $\theta_k$, where $k \in \mathcal{M}$. If we have a number of special nodes $M << N$, there is a method for constructing the circuit only with $M+1$ phase rotation gates. First, we apply a phase rotation of $\theta$ to the second register, with no control by the first register. Thus, a global APR phase $\theta$ is applied for all nodes. After that, we apply a phase rotation for each of the $M$ special nodes, controlled by the first register being in the corresponding state. Let $\theta_k$ be the phase we want to apply to node $k$, which is an special node. Since we have already applied a phase $\theta$ to that node, the particular phase rotation for that node must now apply $\theta_k - \theta$, so that it deletes the previous rotation with $\theta$. In Figure \ref{F:circuit-special-apr} we show an example where node $N-1$ is the only special node. Note that in general each of the special nodes can have a different local APR phase.

\subsection{Adding \link phases}

We have seen that after diagonalizing the reflection operator $R$, the diagonal operator $D$, even with APR phases, does not depend on the vectors $\left|\psi_i\right>$. Thus, it will not depend on the \link phases $\varphi_{ij}$. In this case, the modification must be done to the update operator $\update$. The diagonalization process is the same as before, but using the modified update operator $\update(\varphi)$, such that $\update(\varphi)\left|i\right>_1\left|0\right>_2 = \left|\psi_i(\varphi)\right>$.

Provided a circuit implementation of the update operator $\update$, a naive method to construct the operator $\update(\phimatrix)$ would be to first apply $\update$, and then apply a controlled-phase gate (see Figure \ref{F:phase_gate}) for each computational basis state whose \link phase $\varphi_{ij}$ is different to $0$. In general we would have $N^2$ \link phases, so this method would be inefficient unless there are very few \link phases different to $0$, or they are distributed in a manner that several phases with the same value can be applied using a common controlled gate.

Another approach consists of constructing a new circuit for the modified update operator based on the standard one. The same as with the transition probabilities, if the \link phases matrix $\varphi$ is sparse or the distribution of phases in the graph follows some symmetry patterns, we could construct an efficient operator $\update(\varphi)$. An example of update operator with phases distributed with certain symmetry in cycles will be constructed in section \ref{sec:double-coin}.

\section{Equivalence with the coined quantum walk}\label{sec:Coin}

It is known that given a standard Szegedy's quantum walk, it can always be cast into the coined model, but only a restricted set of coins can be cast into Szegedy's model \cite{Sandbichler-master}. In this section we will review the coined quantum walk formulation and the equivalence with Szegedy's quantum walk, in order to show how the phase extensions allows the \twnan model to host a wider set of equivalent coins.

\subsection{Coined quantum walk formulation}

\begin{figure}[hbpt]
	\centering
	\subfigure[]{\includegraphics[scale=0.6]{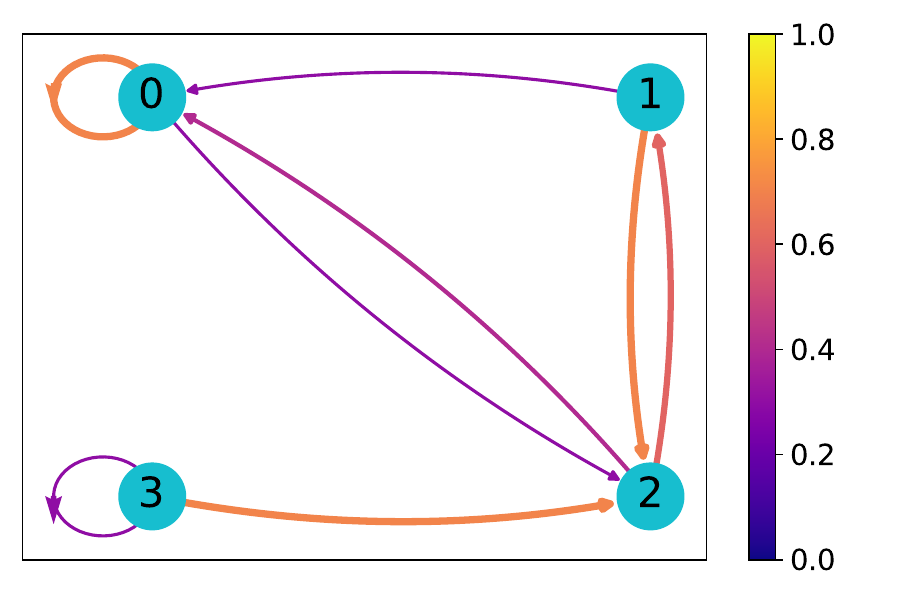}\label{F:weighted_graph}}
	\subfigure[]{\includegraphics[scale=0.56]{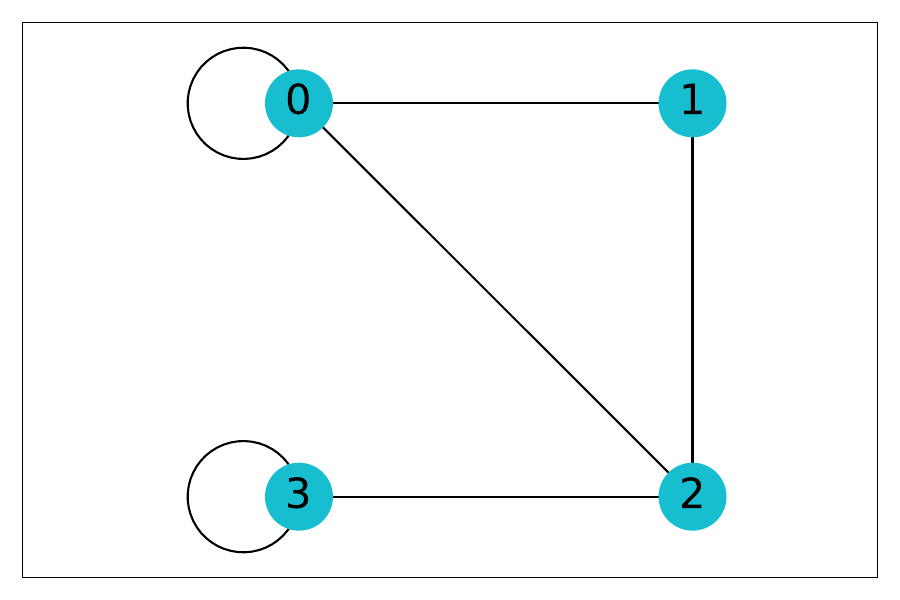}\label{F:undirected_graph}}
	\caption{a) Weighted graph with $N=4$ nodes whose transition matrix is given in \eqref{G4}. The weights of the directed edges are represented by the colormap, and are proportional to the width of the edges. Missing edges have a null transition probability. b) Undirected graph with $N=4$ nodes representing the backbone of the weighted graph in a). Te adjacency matrix is in \eqref{A4}. Note that whereas the weighted graph is asymmetric, the underlying undirected graph is symmetric. For example, in the weighted graph there is a directed edge from node $1$ to node $0$, but there is no directed edge from $0$ to $1$. This asymmetry is not reflected in the undirected graph, where a symmetric edge is present. The graphs have been plotted using the python library NetworkX \cite{NetworkX}.}
	\label{F:...}
\end{figure}

Whereas Szegedy's quantum walk takes place on weighted graphs, coined quantum walks are defined on undirected graphs. In Figure \ref{F:weighted_graph} we show an example of a weighted graph with four nodes whose associated transition matrix is
\begin{equation}\label{G4}
G = \left(\begin{array}{cccc}
	0.7 & 0.3 & 0.4 & 0 \\
	0 & 0 & 0.6 & 0 \\
	0.3 & 0.7 & 0 & 0.7 \\
	0 & 0 & 0 & 0.3
\end{array}\right).
\end{equation}
The underlying undirected graph corresponds to the backbone of the graph, as it is shown in Figure \ref{F:undirected_graph}. Whereas the weighted graph is formed by directed edges with an associated transition probability, the undirected graph is formed by undirected edges, so they do not have an arrow indicating the direction. Moreover, note that the underlying undirected graph has an edge if there is a directed edge in the weighted graph regardless of the direction. For example, between nodes 0 and 1 there is a directed edge from node 1 to node 0, but there is no directed edge from 0 to 1 since the corresponding transition probability is $G_{10} = 0$. However, this asymmetry is not reflected in the undirected graph, where an undirected edge connects nodes 0 and 1.

The adjacency matrix $A$ of an undirected matrix is a symmetric matrix such that $A_{ij}=1$ if there is an edge between nodes $i$ and $j$. Otherwise $A_{ij} = 0$. Let us denote $G^S$ as the result of symmetrizing the transition matrix $G$. Then, $A$ has non-null elements in the same positions as $G^S$. In this case the adjacency matrix is
\begin{equation}\label{A4}
A = 
\left(\begin{array}{cccc}
	1 & 1 & 1 & 0 \\
	1 & 0 & 1 & 0 \\
	1 & 1 & 0 & 1 \\
	0 & 0 & 1 & 1
\end{array}\right).
\end{equation}

Although self-loops are not usually considered, for the sake of establishing a relationship with Szegedy's quantum walk we also need to consider self-loops. Thus, we provide a construction of coined quantum walks on arbitrary graphs \cite{Portugal} considering also self-loops. This construction uses the arc notation \cite{Arc_notation}, so that the computational basis is formed by states representing directed edges of the graph. For undirected graphs we consider that each edge between two different nodes has associated two directed edges, whereas a self-loop only has a single directed edge. Thus, the Hilbert space where the coined quantum walk takes place is
\begin{equation}
	\mathcal{H}_C = \text{span}\lbrace{\left|(i,j)\right> \ : \ A_{ij} = 1\rbrace},
\end{equation}
where $(i,j)$ represents the directed edge pointing from node $i$ to node $j$. The dimension is $2|E_{N}| + |E_{L}|$, where $E_{N}$ is the set of undirected edges between different nodes, and $E_L$ the set of self-loops. For the example of Figure \ref{F:undirected_graph} $E_N = 4$ and $E_L = 2$. Note that due to the symmetry of the undirected graph, if $\left|(i,j)\right>$ is in the Hilbert space, then $\left|(j,i)\right>$ is too.

The unitary evolution operator $U_c$ of the coined quantum walk is defined as follows:
\begin{equation}
	U_c := S_cC,
\end{equation}
where $S_c$ is a shift operator and $C$ the coin operator. For a general coined quantum walk on an undirected graph, the flip-flop shift operator is usually used. Its action is defined by \cite{Flip-flop}:
\begin{equation}\label{flip-flop}
	S_c\left|(i,j)\right> = \left|(j,i)\right>.
\end{equation}
There are other possibilities for the shift operator, as the moving shift operator. However, in the end both quantizations are equivalent via a redefinition of the coin \cite{Portugal}.

The coin operator $C$ is expressed as:
\begin{equation}
	C = \bigoplus_{i=0}^{N-1} C_i,
\end{equation}
so that in general there is a different coin $C_i$ associated to each of the $N$ nodes of the graph. Thus, depending on the node where the walker is, which is represented by the tail of the directed edge, the coin acts differently. Moreover, in general each $C_i$ matrix has a different dimension. Each $C_i$ operator is a $d_i$-dimensional unitary operator, where $d_i$ is the degree of node $i$, and acts in the subspace
\begin{equation}
	\mathcal{H}_{C}^i = \text{span}\lbrace{\left|(i,k)\right> \ : \ A_{ik} = 1\rbrace},
\end{equation}
which is formed by the $d_i$ directed edges departing from node $i$. Usually the Grover diffusion operator \cite{Grover} is used as coin \cite{Portugal}, so that the matrix elements are
\begin{equation}
(C_i)_{ab} = \frac{2}{d_i} - \delta_{ab}.
\end{equation}

\subsection{Relation between coined and Szegedy's quantum walks}

\Apriori we cannot establish a relationship between coined quantum walks and Szegedy's model because both Hilbert spaces have different dimensions. Moreover, the coined space $\mathcal{H}_C$ cannot be expressed as a tensor product since each node has in general a different degree. We can augment the coined space into an $N^2$-dimensional space considering also the edges $(i,j)$ that are not present in the undirected graph. We call $\mathcal{H}_C^A$ to this augmented space, and $\mathcal{H_C}$ is a subspace where the walk takes place. The augmented space now is isomorphic to the Hilbert space of Szegedy's quantum walk $\mathcal{H_S}$ in \eqref{hilbert_szegedy}. Let $\mathcal{F}$ be the isomorphism between both Hilbert spaces:
\begin{equation}
	\mathcal{F}: \mathcal{H}^A_C \rightarrow \mathcal{H}_S.
\end{equation}
Using it on $\mathcal{H_C}$ we can find a reduced subspace of $\mathcal{H}_S$ where the equivalent Szegedy's quantum walk should take place. Let us denote it as $\mathcal{H}^R_S$, so that $\mathcal{F}: \mathcal{H}_C \rightarrow \mathcal{H}^R_S$ and
\begin{equation}
	\mathcal{H}_S^R = \text{span}\lbrace{\left|i\right>_1\left|j\right>_2 : A_{ij} = 1\rbrace}.
\end{equation}

Now we need to define an augmented $N^2$-dimensional coined walk operator compatible with the augmented space. We define it directly acting on $\mathcal{H}_S$ as
\begin{equation}
U^A_c = S_c^AC^A.
\end{equation}
For this operator to be equivalent to the coined walk in $\mathcal{H_C}$, it must have the same action as $U_c$ in the reduced subspace $\mathcal{H}^R_S$ after the application of the isomorphism, and leaving it invariant. For the shift operator we can take $S^A_c = S_w$. From \eqref{flip-flop} it is trivial that the swap operator $S_w$ acts equivalently to $S_c$ in the reduced subspace. Moreover, $\mathcal{H}_S^R$ is trivially invariant under $S_w$ due to the symmetry of the undirected adjacency matrix $A$.

With regard to the coin, we can define a coin operator in $\mathcal{H}_S$ as:
\begin{equation}\label{augmented_coin}
	C^A = \sum_{i=0}^{N-1} \left|i\right>_1\left<i\right| \otimes C^A_i,
\end{equation}
where now $C^A_i$ is a $N$-dimensional unitary operator that acts non-trivially in
\begin{equation}
	\mathcal{H}^i_S = \text{span}\lbrace{\left|k\right>_2 \ : \ A_{ik} = 1\rbrace},
\end{equation}
conditioned by the first register being in state $\left|i\right>_1$. In order to the coin $C^A$ be equivalent to the coin $C$, the action in the reduced subspace must be provided by the isomorphism $\mathcal{F}$ as
\begin{equation}
	C^A\left|i\right>_1\left|j\right>_2 := 
	\left\lbrace\begin{array}{c}
		\displaystyle \mathcal{F}(C\left|(i,j)\right>) \ \ \ \text{if} \ \left|i\right>_1\left|j\right>_2 \in \mathcal{H}^R_S,\\
		\\
		\displaystyle -\left|i\right>_1\left|j\right>_2 \ \ \  \text{if} \ \left|i\right>_1\left|j\right>_2 \in \left(\mathcal{H}^{R}_S\right)^\perp.
	\end{array}
	\right.
\end{equation}
We need the reduced subspace to be invariant under the action of $C^A$. The action on the states that are perpendicular to $\mathcal{H}^R_S$ is irrelevant as long as it does not mix the subspaces. Thus, we have freedom defining the action on the orthogonal complement. For the sake of establishing later an equivalence with Szegedy's quantum walk, we define this action as the $-\mathbbm{1}$ operator.

Since we have defined $S_c^A = S_w$, the equivalence with the \twnan Szegedy's quantum walk needs the operator $R(\vec{\theta},\varphi)$ in \eqref{general_R} to be the coin operator $C^A$. For a state $\left|i\right>_1\left|j\right>_2$ in the orthogonal complement of $\mathcal{H}^R_S$ we have $A_{ij} = 0$, which implies that in the transition matrix $G_{ij}=G_{ji} = 0$. Thus, this state is perpendicular to all the $\left|\psi_i(\varphi)\right>$ states in \eqref{psi_i_extended} and the action of $R(\vec{\theta},\varphi)$ is just $-\mathbbm{1}$. Thus, this subspace is invariant under $R(\vec{\theta},\varphi)$, and due to unitarity $\mathcal{H}_S^R$ also is. Therefore, we are closer to establish $C^A = R(\vec{\theta},\varphi)$, and $U_c^A = U_s(\vec{\theta},\varphi)$.

The last step is to find the expression of the individual coins $C^A_i$ in \eqref{augmented_coin}. Let us rewrite the $\left|\psi_i(\varphi)\right>$ states as:
\begin{equation}
	\left|\psi_i(\varphi)\right> = \left|i\right>_1 \otimes \left|\omega_i(\varphi)\right>_2,
\end{equation}
where
\begin{equation}\label{c_omega}
	\left|\omega_i(\varphi)\right>_2 = \sum_{k=0}^{N-1} e^{\ci\varphi_{ik}}\sqrt{G_{ki}}\left|k\right>_2.
\end{equation}
Taking into account that $\mathbbm{1} = \mathbbm{1}_1 \otimes \mathbbm{1}_2$, the completeness relation of the identity for the first register, and substituting the expression for the $\left|\psi_i(\varphi)\right>$ states in \eqref{general_R}, we have:
\begin{equation}\label{R_coin}
	R(\vec{\theta},\varphi) = \sum_{i=0}^{N-1} \left|i\right>_1\left<i\right| \otimes \left[(1-e^{\ci\theta_i})\left|\omega_i(\varphi)\right>_2\left<\omega_i(\varphi)\right| - \mathbbm{1}_2\right].
\end{equation}
Looking at the expression for the coin $C^A$ in \eqref{augmented_coin}, we can identify the individual coins $C^A_i$ with the phase rotations in the second register:
\begin{equation}\label{coin_reflection}
	C^A_i = (1-e^{\ci\theta_i})\left|\omega_i(\varphi)\right>_2\left<\omega_i(\varphi)\right| - \mathbbm{1}_2.
\end{equation}
When the augmented coins can be expressed in this form, then the coined quantum walk is equivalent to Szegedy's quantum walk. The coins codify the transition probabilities $G_{ji}$ and also the extended phases, and we have finally
\begin{equation}
	\mathcal{F}(U_c\left|(i,j)\right>) = U_s(\vec{\theta},\varphi)\left|i\right>_1\left|j\right>_2.
\end{equation}
Note that $\mathcal{H}^R_S$ is invariant under Szegedy's quantum walk and it is actually where any Szegedy's quantum walk takes place for a weighted graph, since it contains all the directed edges with non-null probability and their swapped versions, which appear due to the swap operator. Thus, Szegedy's quantum walk is indeed quantized also on the undirected graph. In the case that for an undirected edge one of the two transition probabilities is null, this is taken into account by the coin. Nevertheless, this ghost directed edge plays also a role in the quantum state and cannot be removed.

The question that remains is what are the conditions that must be satisfied so that a coin can be expressed as in \eqref{coin_reflection}. In the case that we are provided with a Szegedy's quantum walk, we can always define the coins that way, so that all Szegedy's quantum walk that come from the quantization of a classical Markov chain with transition matrix $G$ and extended phases $\vec{\theta}$ and $\varphi$ can be cast into the coined model. In the case that the weighted graph is obtained normalizing the columns of the adjacency matrix of an undirected graph, the equivalent coin is the Grover coin \cite{Notes,Wong_1}. Thus, for arbitrary weighted graphs Szegedy's quantum walk is a generalization of Grover's quantum walk \cite{Grover_coin}.

If on the contrary we are provided with a coined quantum walk, only a restricted set of coins satisfy equation \eqref{coin_reflection}. From it we can formulate the following lemmas about the conditions that must be satisfied by a coined walk to be cast into Szegedy's quantum walk. Let us start with the standard model.

\begin{lemma}\label{L:standard-coin}
	Given a coined quantum walk $U_c$ with a set of coin operators $C_i$ of dimension $d_i$, there exists an equivalent standard Szegedy's quantum walk $U_s$ if and only if for each coin $C_i$ there are $d_i-1$ eigenvalues $-1$, and a single eigenvalue $+1$ whose eigenvector has real non-negative amplitudes.
\end{lemma}

In this case the coin must be a reflection around the state $\left|\omega_i\right>$, so that $\left|\omega_i\right>$ is an eigenstate with eigenvalue $+1$, and the rest of eigenvalues are $-1$. Moreover, all the amplitudes of $\left|\omega_i\right>$ are real positive or zero. This condition for the coins to be able to be cast into the standard Szegedy's model is in concordance with the conditions found in the literature \cite{Sandbichler-master}. Note that since eigenvectors that differ in a global phase are equivalent, the actual condition for the eigenvector is that there are no relative phases between the amplitudes.

\begin{lemma}\label{L:vector-coin}
	Given a coined quantum walk $U_c$ with a set of coin operators $C_i$ of dimension $d_i$, there exists an equivalent \linkphased Szegedy's quantum walk $U_s(\varphi)$ if and only if for each coin $C_i$ there are $d_i-1$ eigenvalues $-1$, and a single eigenvalue $+1$.
\end{lemma}

In this case the coin must be a reflection around the state $\left|\omega_i(\varphi)\right>$. Again all the eigenvalues must be $-1$ except for the eigenstate $\left|\omega_i(\varphi)\right>$, which is $+1$. However, thanks to the \link phases this eigenstate can have any complex amplitudes. So that this model can host a bigger set of coins.

\begin{lemma}\label{L:apr-coin}
	Given a coined quantum walk $U_c$ with a set of coin operators $C_i$ of dimension $d_i$, there exists an equivalent \vertexphased Szegedy's quantum walk $U_s(\vec{\theta})$ if and only if for each coin $C_i$ there are $d_i-1$ eigenvalues $-1$, and a single eigenvalue $-e^{\ci\theta_i}$ whose eigenvector has real non-negative amplitudes.
\end{lemma}

If we add APR phases, the coin becomes a phase rotation operator. Thus, $\left|\omega_i\right>$ is an eigenstate with eigenvalue $-e^{\ci\theta_i}$, and the rest of eigenvalues are $-1$. Note that if we would only have a model with global APR, all the coins should have the same eigenvalue different to $-1$. However, thanks to the local APR phases introduced in this work, each node can have a coin with different eigenvalues.

\begin{lemma}\label{L:generalized-coin}
Given a coined quantum walk $U_c$ with a set of coin operators $C_i$ of dimension $d_i$, there exists an equivalent \twnan Szegedy's quantum walk $U_s(\vec{\theta},\varphi)$ if and only if for each coin $C_i$ there are $d_i-1$ eigenvalues $-1$, and a single eigenvalue $-e^{\ci\theta_i}$.
\end{lemma}

Considering both \link phases and local APR phases the eigenstate $\left|\omega_i(\varphi)\right>$ can have any complex amplitude at the same time that the eigenvalue is arbitrary. Therefore the set of coins that can be cast is maximal, and encompasses the previous cases.

All these equivalences have been established between the coined walk and the single step Szegedy's operator $U_s$. If we can cast a coined walk into Szegedy's in this case, then trivially it can also be cast into Szegedy's model considering the original double operator $W_s$, so that one step of Szegedy's quantum walk would be equivalent to two steps of the coined walk, being the equivalent operator $U^2_c$. Moreover, there can be cases where a coin cannot be cast into Szegedy's model considering the single step operator $U_s$, but it can be cast if we consider the double step operator $W_s$ instead. We will show an example for the $-\mathbbm{1}$ coin in section \ref{sec:search}.

\begin{figure}[htpb]
	\centering
	\subfigure[]{\includegraphics[scale=0.56]{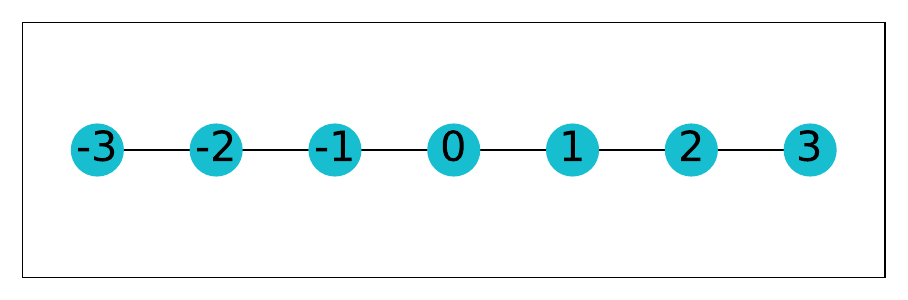}\label{F:undirected_line}}
	\subfigure[]{\includegraphics[scale=0.6]{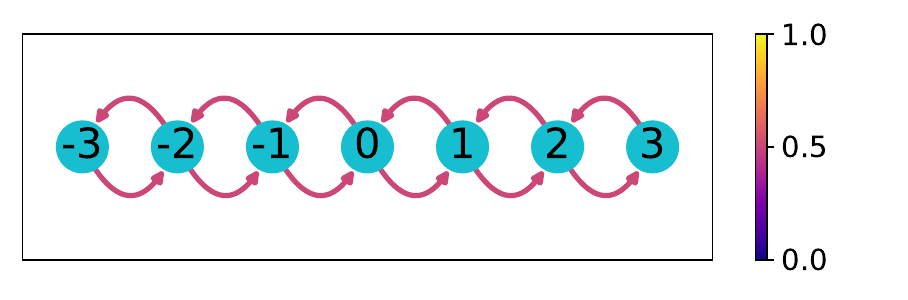}\label{F:line_X}}
	\subfigure[]{\includegraphics[scale=0.6]{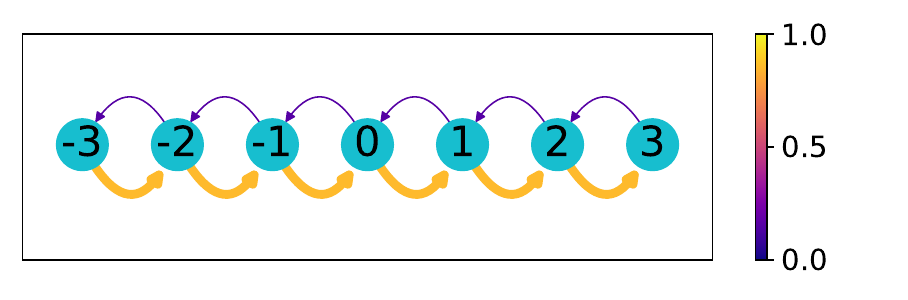}\label{F:line_H}}
	\caption{a) Undirected graph for an infinite 1D line. For the sake of simplicity only seven nodes have been represented. b) Weighted graph obtained by normalizing the adjacency matrix of the undirected graph in a), so that all directed edges have the same transition probability. c) Weighted graph associated to the coined walk with the Hadamard coin, where the probability to the right is bigger than to the left. The weights of the directed edges are represented by the colormap, and are proportional to the width of the edges. The graphs have been plotted using the python library NetworkX \cite{NetworkX}.}
	\label{F:...}
\end{figure}

\section{Example on the line}\label{sec:Line}

So far we have obtained the conditions for establishing an equivalence between Szegedy's model and coined quantum walks. In this section we will show how each model can be cast into the other using examples on the 1D line. The undirected graph that shows the backbone of the line is shown in Figure \ref{F:undirected_line}. The set of nodes ranges from $-\infty$ to $\infty$, although we only show seven nodes for the sake of simplicity.

\subsection{Standard Szegedy's model and coined quantum walk}

\begin{figure*}
	\centering
	\makebox[10pt][c]{
		\includegraphics[scale=0.6]{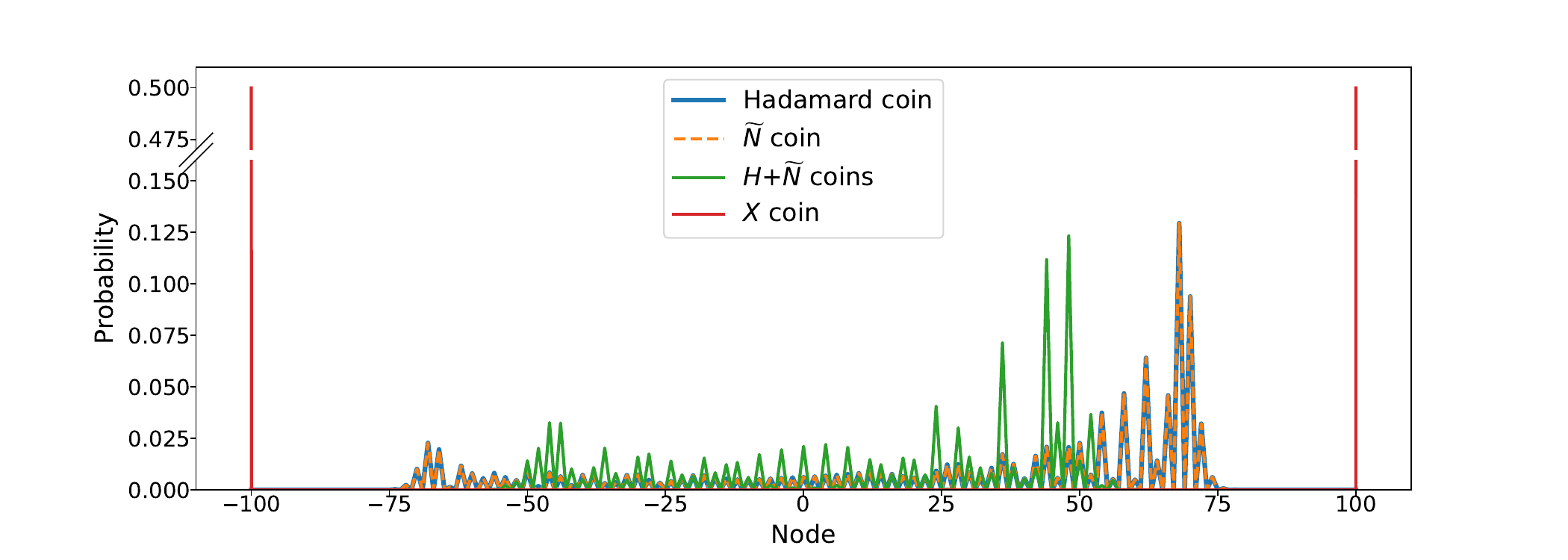}
	}
	\caption{Probability distribution results of the coined quantum walk simulations on a 1D line using the $X$ coin (red), the Hadamard coin (blue), the $\coin$ coin (orange), and the coined walk with the Hadamard coin for even nodes and the $\coin$ coin for odd nodes (green). Note that for the Hadamard and $\coin$ coins the curves overlap. The initial state is $\left|\psi_0\right> = \left(\left|0\right>_1\left|1\right>_2 + \left|0\right>_1\left|-1\right>_2\right)/\sqrt{2}$ and the unitary evolution has been applied $100$ time steps.}
	\label{F:all_coin}
\end{figure*}

The naive transition matrix used for standard Szegedy's quantization is obtained by normalizing the adjacency matrix, so that a walker in node $i$ has a probability of $1/d_i$ for jumping to each of the neighbor nodes. In this case the walker has a probability of one half for jumping either to the right or to the left. The associated weighted graph is shown in Figure \ref{F:line_X}, and the transition matrix is
\begin{equation}\label{undirected_G}
	\left(G_u\right)_{ji} = \frac{1}{2}\delta_{j-1,i} + \frac{1}{2}\delta_{j+1,i},
\end{equation}
where the subindex $u$ makes reference to the fact that it is obtained from the undirected graph. Since the degree of each node is $d_i = 2$, all coins $C_i$ are going to be $2$-dimensional matrices. Moreover, all coins will be the same. By convention the 2D basis is ordered so that the first element corresponds to the directed edge pointing to the right, and the second one pointing to the left. Since there are no extended phases, the coins correspond to reflections around the state $\left|\omega_i\right> = (1,1)^T/\sqrt{2}$, which is obtained taking the square root of the columns of $G_u$ and expressing it directly in the 2D basis of the coin. The rest of elements in the augmented space are always null. This is an eigenvector with eigenvalue $+1$, and the other eigenvalue must be $-1$. Thus, it trivially corresponds to the Pauli $X$ operator, which is the Grover coin in 2D.

To compare different quantum walks, we need an initial state for the simulation. We take as initial state $\left|\psi_0\right> = \left(\left|0\right>_1\left|1\right>_2 + \left|0\right>_1\left|-1\right>_2\right)/\sqrt{2}$, which represents the walker at node $0$. The probability distribution after $100$ time steps is shown in Figure \ref{F:all_coin} in red. The probability distribution moves ballistically, so that there are no probabilities at intermediate nodes, and the walker reach nodes $100$ and $-100$ with the same probability.

A different quantization of the line walk is usually done by means of the Hadamard coin \cite{Coined-general-graph}. Since this coin puts in an equal superposition the computational basis of the coin register in 2D, it was thought to be a sensible quantization of the classical walk on the undirected graph with $G_u$. However, when casting it into a Szegedy's quantum walk we obtain an unbiased transition matrix which does not corresponds to the classical walk on the undirected line \cite{Sandbichler-master}. The spectrum of $H$ is $\sigma(H) = \lbrace{+1,-1\rbrace}$, and the eigenvector for eigenvalue $+1$ is $(h_R,h_L)^T$, where
\begin{equation}\label{h_R}
h_R = \frac{1}{\sqrt{4-2\sqrt{2}}},
\end{equation}
\begin{equation}\label{h_L}
h_L = \frac{\sqrt{2}-1}{\sqrt{4-2\sqrt{2}}}.
\end{equation}
Thus, by Lemma \ref{L:standard-coin} we can cast it into the standard Szegedy's model. The transition probabilities are obtained by the square modulus of the amplitudes $h_R$ and $h_L$, so that the walker has a probability of approximately $0.85$ of jumping to the right and approximately $0.15$ of jumping to the left. The transition matrix is therefore
\begin{equation}
	\left(G_H\right)_{ji} = h_R^2\delta_{j-1,i} + h_L^2\delta_{j+1,i}.
\end{equation}
The associated weighted graph is shown in Figure \ref{F:line_H}, where we observe that the walk is biased to the right. The results of the simulation for $100$ time steps using the same initial state as before are shown in Figure \ref{F:all_coin} in blue. In this case the walk is slower, not reaching nodes $\pm 100$. Moreover, there are probabilities at intermediate nodes, and the distribution is biased to the right.

Note that despite the fact that the Hadamard coined walk and the unbiased Szegedy's walk are two quantizations based on the undirected graph, they are not equivalent. The equivalence between both models just means that given a transition matrix we can find a set of coins that reproduce that particular Szegedy's walk. However, different coins produce different quantizations that may be equivalent to quite different weighted graphs, despite being devised from the same classical walk as in the case of the Hadamard coin.

\subsection{\Twnan Szegedy's model and coined quantum walk}

Now let us show a different coin that cannot be cast into the standard Szegedy's model. We have constructed the following coin:
\begin{equation}
	\coin = -\frac{1+\ci}{2} 
	\left(\begin{array}{cc}
		1 & 1 \\
		-1 & 1
	\end{array}\right).
\end{equation}
The spectrum is $\sigma(\coin) = \lbrace{-\ci,-1\rbrace}$ and the eigenvector for eigenvalue $-\ci$ is $(1,\ci)^T/\sqrt{2}$. Thus, by Lemma \ref{L:standard-coin} it cannot be cast into the standard model, but by Lemma \ref{L:generalized-coin} it can be cast into the \twnan model. In this case the eigenvalue different to $-1$ was $-e^{\ci\theta_i}$, so that the APR phase for this coin is $\theta_i = \pi/2$. Again the transition probabilities are obtained taking the square modulus of the amplitudes of the eigenvector. In this case we obtain $1/2$ for both directions, so that the transition matrix $G_{\coin}$ associated to this coin is the same as for the undirected graph $G_u$ in \eqref{undirected_G}. Since the amplitudes are complex numbers, we also need \link phases. For directed edges pointing to the right the amplitude is a real positive number and the \link phase is $\varphi_{i,i+1} = 0$. However, for directed edges pointing to the left the \link phase is $\varphi_{i,i-1} = \pi/2$. Thus, the \link phases matrix for this coin is
\begin{equation}
	\left(\varphi_{\coin}\right)_{ij} = \frac{\pi}{2}\delta_{i,j+1}.
\end{equation}

The results of the simulation using this coin for all nodes are shown in Figure \ref{F:all_coin} in orange. The probability distribution is the same as for the Hadamard coin. This is surprising since for the $\coin$ coin the associated weighted graph is unbiased, being the same as for the $X$ coin. Thus, the extended phases play an important role in the walk, being able to modify somehow the transition probabilities.

Since both the Hadamard coin and the $\coin$ coin produce the same results when they are used on all nodes, we want to examine what happens if we apply both at the same time on different nodes. We have quantized the walk on the line using the Hadamard coin for even nodes, and the $\coin$ coin for odd nodes. Taking into account our convention for the transition matrix indexes, it is constructed taking the even columns of $G_H$ and the odd columns of $G_{\coin}$. For the \link phases matrix $\varphi$ we take the odd rows of $\varphi_{\coin}$, and the even rows are null since the Hadamard coin has no \link phases. The local APR phase for even nodes is the standard one $\theta_i = \pi$ and for odd nodes is $\theta_i = \pi/2$.

The results of the simulation are shown in Figure \ref{F:all_coin} in green. We observe that despite the fact that $H$ and $\coin$ produce the same results when they are global coins, the results are different when they are mixed. This explicitly shows that both coins are not actually equivalent, and that the \twnan Szegedy's model opens up a wide range of possibilities for quantizing classical Markov chains.

\subsection{Quantum circuit for the $H+\coin$ coined quantum walk}\label{sec:double-coin}

Before finishing this section, we want to provide a quantum circuit construction for the \twnan Szegedy's quantum walk equivalent to the coined walk that uses both coins at the same time. A quantum walk in the infinite 1D line can be simulated in a finite 1D cycle as long as the time steps do not surpass a threshold from which the wavefront interferes with itself when it completes a turn on the cycle. For a graph with $N=2^n$ nodes, we need $n$ qubits for each register.


\begin{figure}[hbpt]
	\centering
	\includegraphics[scale=1]{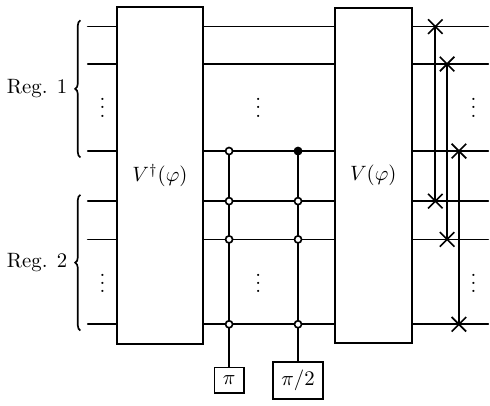}
	\caption{Quantum circuit decomposition of the \twnan single Szegedy unitary evolution operator $U_s=S_wR(\vec{\theta},\varphi)$ for the coined walk with the coins $H$ and $\coin$. Each register has $n$ qubits for a cycle with $N=2^n$ nodes. The last qubit of the first register, which determines the parity of the state, controls the application of a local APR phase $\pi$ for even nodes, and $\pi/2$ for odd nodes.}
	\label{F:circuit-double-coin}
\end{figure}


\begin{figure*}
	\centering
	\includegraphics[scale=1]{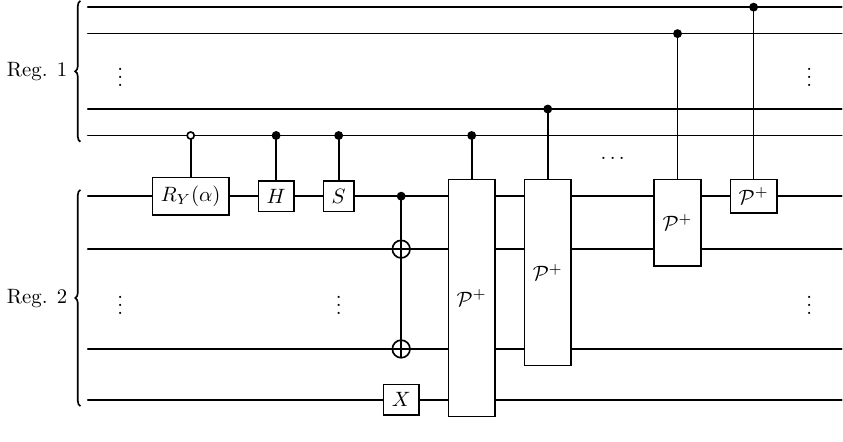}
	\caption{Quantum circuit for the update operator $\update(\varphi)$ of the \twnan Szegedy's walk equivalent to the coined walk with the $H$ coin for even nodes and the $\coin$ coin for odd nodes. The rotation matrix is $R_Y(\alpha) = \exp{-\ci\alpha Y/2}$, with $Y$ being a Pauli matrix, and the value of the rotation angle is $\alpha=\pi/4$.}
	\label{F:circuit-double-coin-update}
\end{figure*}

In this case the local APR phases are distributed with a symmetric pattern, so that the diagonal operator $D(\vec{\theta})$ can be implemented efficiently. The last qubit of the first register indicates the parity of the node. If this control qubit is in state $\left|0\right>$, for even nodes we apply the phase rotation with $\theta = \pi$ on the second register. If on the contrary it is on state $\left|1\right>$, for odd nodes the phase rotation with $\theta = \pi/2$ is applied instead. Thus, the quantum circuit can be decomposed as shown in Figure \ref{F:circuit-double-coin}.

For the update operator $\update(\phimatrix)$, in Figure \ref{F:circuit-double-coin-update} we have constructed a circuit based on the update operator of the standard cyclic graph \cite{Q_circuits}. The update operator must act as $\update(\phimatrix) \left|i\right>_1\left|0\right>_2 = \left|i\right>_1\left|\omega_i(\varphi)\right>_2$, where
\begin{equation}
	\left|\omega_i(\varphi)\right>_2 = 
	\left\lbrace\begin{array}{c}
		\displaystyle h_R\left|i+1\right>_2 + h_L\left|i-1\right>_2 \ \ \ \text{if $i$ is even}, \\
		\\
		\displaystyle \frac{1}{\sqrt{2}}\left|i+1\right>_2 + \frac{\ci}{\sqrt{2}}\left|i-1\right>_2 \ \ \text{if $i$ is odd},
	\end{array}\right.
\end{equation}\\
where $h_R$ and $h_L$ are given in \eqref{h_R} and \eqref{h_L} respectively. Since we are considering a finite graph, the additions and subtractions are performed modulo $N$.

Let $i$ be an even node. Then, the last qubit of the first register is in state $\left|0\right>$, and the $R_Y$ gate is applied to the first qubit of the second register, letting it in state $h_R\left|0\right>_2 + h_L\left|1\right>_2$. The following action of the CNOT gates and the last $X$ gate let the second register in
\begin{equation}\label{partial_state_even}
h_R\left|1\right>_2 + h_L\left|N-1\right>_2.
\end{equation}
The permutation operator $\adder$ acting on a $m$-qubit system transforms the computational basis states $\left|x\right>$ into $\left|x+1 \ \text{mod} \ 2^m\right>$. This can be implemented efficiently with multi-controlled-NOT gates. The join action of the $\adder$ operators controlled by the first register this way corresponds to another permutation that transforms the computational basis states $\left|i\right>_1\left|x\right>_2$ into $\left|i\right>_1\left|x+i \ \text{mod} \ N\right>_2$ \cite{Q_circuits}. Therefore, when they are applied on the state in \eqref{partial_state_even}, the second register ends up in $\left|\omega_i(\varphi)\right>_2$.

We can do an analogous reasoning for $i$ being an odd node. In this case the last qubit of the first register is in state $\left|1\right>$, so that the $H$ and $S$ gates are applied to the first qubit of the second register, letting it in state $\left(\left|0\right> + \ci\left|1\right>\right)/\sqrt{2}$. After the CNOT gates and the last $X$ gate the second register ends up in
\begin{equation}
\frac{1}{\sqrt{2}}\left|1\right>_2 + \frac{\ci}{\sqrt{2}}\left|N-1\right>_2,
\end{equation}
and the controlled $\adder$ operators permute the state into $\left|\omega_i(\varphi)\right>_2$. Thus, we have proved that the circuit in Figure \ref{F:circuit-double-coin-update} effectively performs the action of the update operator $\update(\varphi)$.

\section{Marking nodes with APR}\label{sec:Marking}

Suppose there is a special set of nodes in a graph that we want to mark somehow so that the quantum evolution treats them in a different manner. We can mark them with a different local APR phase. A construction based on Figure \ref{F:circuit-special-apr} would require the knowledge about who are these special nodes. However, we usually do not know them, and they are marked by a black-box function when they satisfy some conditions.


\begin{figure}[hbpt]
	\centering
	\includegraphics[scale=1]{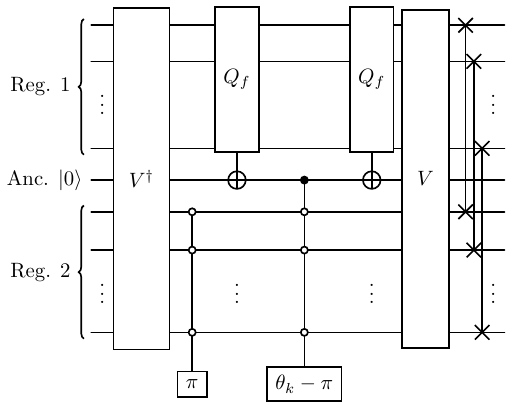}
	\caption{Quantum circuit decomposition for marking nodes with local APR phases in a standard Szegedy's quantum walk. The first oracle turns the ancilla qubit into $\left|1\right>$ if the first register represents a marked nodes. Then, the ancilla applies a phase rotation on the second register deleting the global APR phase $\pi$ and letting the local APR phase $\theta_k$. Finally, the second oracle uncomputes the ancilla so that it returns to its original state and can be traced out.}
	\label{F:circuit-oracle}
\end{figure}

Let $f(x)$ be a classical function that decides if a particular node $x$ satisfies some conditions, and let $\mathcal{M}$ be the set of nodes that satisfy them. Then
\begin{equation}
	f(x) := 
	\left\lbrace\begin{array}{c}
		1 \ \ \ \text{if} \ x \in \mathcal{M},\\
		\\
		0 \ \ \ \text{otherwise}.
	\end{array}
	\right.
\end{equation}
Provided a classical circuit can be constructed for this function $f(x)$, it can be translated into quantum gates so that we can construct a quantum circuit for an oracle $Q_f$. This operator takes a quantum register with the information about the node $x$ and adds the result of $f(x)$ in an ancilla qubit \cite{Nielsen,MA_review_modern_physics}, such that
\begin{equation}
	Q_f\left|x\right>\left|y\right> = \left|x\right>\left|y \oplus f(x)\right>.
\end{equation}
When the ancilla qubit starts in state $\left|0\right>$ it just stores the result of $f(x)$.

For the sake of simplicity let us consider that we want to mark some nodes in a standard Szegedy's walk so that they evolve with a different local APR phase when they satisfy the conditions of an oracle $Q_f$. We can construct a quantum circuit leveraging the update operator $\update$ of the standard walk, as shown in Figure \ref{F:circuit-oracle}. In this case we make use of an ancilla qubit that takes the value of $f(x)$ to later control the phase rotation that deletes the global phase $\pi$ and let the local APR phase $\theta_k$ for marked nodes. Thus, a marked node $k \in \mathcal{M}$ evolves with the local APR phase $\theta_k$, instead of the standard phase $\pi$. The second oracle is used to uncompute the action of the first one, so that the ancilla qubit returns to its original state and can be traced out. Note that although we have used only an oracle for marking nodes, we can use multiple oracles to mark different sets of nodes with different phases, so that they evolve differently depending on the different conditions imposed by the oracles.

\subsection{Searching marked nodes}\label{sec:search}

The \link phases have been previously used to search for marked arcs on graphs \cite{Signed_search}. In this section we show that an important use case of the local APR phases is also the search problem. Nevertheless, in this case we want to use a quantum walk algorithm to search for marked nodes instead of marked arcs.

The first quantum walk search algorithm used the Grover coin for unmarked nodes, and a different coin for marked nodes \cite{QRW_Search}. In particular, it was studied the case of the $-\mathbbm{1}$ coin for marking nodes, finding an algorithm with quadratic speedup with respect to classical ones. This coin cannot be cast into the standard Szegedy's walk since all the eigenvalues are $-1$. However, by Lemma \ref{L:apr-coin} it can be cast into the \vertexphased Szegedy's model using $\theta_k=0$ for marked nodes.

A different approach for quantum walk search algorithms is based on Szegedy's quantum walk with absorbing vertices \cite{Szegedy,Portugal}. Given a transition matrix $G$ for a graph, nodes are marked turning them into sinks, so that they have a self-loop that does not allow the walker to escape once the walker steps on them. Let us denote the modified transition matrix as $G'$. This is obtained deleting the columns of the marked nodes and allocating a diagonal $1$ element, so that
\begin{equation}
	G^{'}_{ji} =
	\left\lbrace\begin{array}{c}
		\displaystyle G_{ji} \ \ \ \text{if} \ i \ \notin \mathcal{M}, \\
		\\
		\displaystyle \delta_{ji} \ \ \ \text{if} \ i \ \in \mathcal{M}.
	\end{array}\right.
\end{equation}

It was previously stated that marking with the $-\mathbbm{1}$ coin is equivalent to marking with absorbing vertices, although the proof was done assuming that the initial state has no amplitudes associated to self-loops \cite{Wong_1}. In this work we will show that both walks are indeed not equivalent for arbitrary states, although they are equivalent when the double step Szegedy operator is considered. Moreover, we make the proof for general graphs, so that the global coin for unmarked nodes does not have to be the Grover coin.

On the one hand, the evolution operator for Szegedy's quantum walk using the $-\mathbbm{1}$ coin for marking nodes is $U_s(\vec{\theta}) = S_wR(\vec{\theta})$ and is obtained setting $\theta_i = \pi$ for unmarked nodes and $\theta_i = 0$ for marked nodes in \eqref{multi-apr}. On the other hand, the evolution operator for Szegedy's quantum walk with absorbing vertices $U_s'$ is obtained from the standard operator in \eqref{U} using the transition matrix $G'$. If we express the reflection operators in a coined form in a similar manner as in equation \eqref{R_coin}, and separate the sums for marked and unmarked nodes, the expressions of these operators are the following:
\begin{widetext}
\begin{equation}
	U_s(\vec{\theta}) = S\sum_{i \notin \mathcal{M}} \left|i\right>_1\left<i\right| \otimes \left(2\left|\omega_i\right>_2\left<\omega_i\right| - \mathbbm{1}_2\right) + S\sum_{i \in \mathcal{M}} \left|i\right>_1\left<i\right| \otimes \left( - \mathbbm{1}_2\right),
\end{equation}
\begin{equation}
	U_s^{'} = S\sum_{i \notin \mathcal{M}} \left|i\right>_1\left<i\right| \otimes \left(2\left|\omega_i\right>_2\left<\omega_i\right| - \mathbbm{1}_2\right) + S\sum_{i \in \mathcal{M}} \left|i\right>_1\left<i\right| \otimes \left(2\left|i\right>_2\left<i\right| - \mathbbm{1}_2\right).
\end{equation}
\end{widetext}

\Apriori both operators are not equivalent, since for marked nodes $U_s(\vec{\theta})$ applies the $-\mathbbm{1}$ coin, whereas $U_s'$ applies a reflection around the self-loop state of the marked node.

We need to calculate how these operators act on the different states of the computational basis in order to examine properly the equivalence. For both operators the first term, corresponding to unmarked nodes, is the same. This is so because the APR phase for unmarked nodes is $\theta_i=\pi$, and the corresponding columns of $G$ and $G'$ are also the same. Thus, the action on a computational basis state $\left|i\right>_1\left|j\right>_2$ with $i \notin \mathcal{M}$ is the same.

Now let us consider a state $\left|i\right>_1\left|j\right>_2$ with $i \in \mathcal{M}$. In this case the action is given by the second terms of the operators. If $i \neq j$, the action is $-S$ in both cases, so it is the same. However, if $i=j$, the action of $U_s(\vec{\theta})$ is $-S$, but the action of $U_s'$ is $+S$, so that $U_s(\vec{\theta})$ acts as $-U_s'$.

Let us define $\mathcal{I}_{ML}$ as the subspace spanned by the marked self-loop states of the computational basis, i.e.
\begin{equation}
	\mathcal{I}_{ML} = \text{span}\left\lbrace\left|i\right>\left|i\right> \ : \ i \in \mathcal{M}\right\rbrace.
\end{equation}
This space is an eigenspace for both operators, with eigenvalue $+1$ for $U_s^{'}$ and $-1$ for $U_s(\vec{\theta})$. Thus, it and its orthogonal complement are invariant subspaces. We can then factorize the action as follows:
\begin{equation}\label{relation_coin_absorbing}
	U_s(\vec{\theta})\left|i\right>\left|j\right> = 
	\left\lbrace\begin{array}{c}
		\displaystyle -U_s^{'}\left|i\right>\left|j\right> \ \ \ \text{if} \left|i\right>\left|j\right> \in \mathcal{I}_{ML},\\
		\\
		\displaystyle \ \ \ U_s^{'}\left|i\right>\left|j\right> \ \ \ \text{if} \left|i\right>\left|j\right> \in \mathcal{I}_{ML}^\dagger.
	\end{array}
	\right.
\end{equation}
Although both operators are not the same, since both subspaces are invariant the action is equivalent when the initial state is in $\mathcal{I}_{ML}^\dagger$. This corresponds to a state that has no amplitudes related to self-loops, so that the previously known equivalence in this case holds \cite{Wong_1}. Nevertheless, note that for general states with self-loops, the only difference will be a relative phase $-1$ in the amplitudes of the self-loops. This relative phase plays no role in the probabilities of measuring the nodes in the computational basis, so that the results would be the same even in this case although the quantum states are not exactly equal. However, this equivalence would not be valid in case we measured in other different basis.

Usually, in quantum walk search algorithms based on absorbing vertices the double Szegedy operator $W_s$ is used. Since the subspaces are invariant, we can take the square of equation \eqref{relation_coin_absorbing}, obtaining that
\begin{equation}
	W_s(\vec{\theta}) = W_s^{'}.
\end{equation}
Thus, for this operator both walks are totally equivalent, taking into account that one Szegedy's step with absorbing vertices corresponds actually to two steps of the coined walk with the $-\mathbbm{1}$ coin. Recall that by Lemma \ref{L:standard-coin} the $-\mathbbm{1}$ coin cannot be cast into a single step of the standard Szegedy's model. However, this Lemma does not apply for the double Szegedy operator, since we have managed to cast this coin into an operator $W_s'$ without phase extensions.

\subsection{Example: complete graph}

A well studied graph for quantum walk search algorithms based on absorbing vertices is the complete graph without loops \cite{Portugal_complete}. The transition matrix is $G_{ji} = (1-\delta_{ji})/(N-1)$. The initial state of the system is constructed from the transition matrix of the original complete graph without marked nodes as
\begin{equation}\label{initial}
	\left|\Psi^{(0)}\right> := \frac{1}{\sqrt{N}} \sum_{i=0}^{N-1} \left|\psi_i\right>.
\end{equation}
In this case the initial state has no self-loop amplitudes, so that the coined walk would be equivalent to the single step operator $U_s'$. Nevertheless, the quantum walk operator is usually $W_s^{'}$. A maximum of probability for measuring one of the marked nodes occurs after a number of time steps given by \cite{Portugal_complete}:
\begin{equation}
	t_{max} = \frac{\pi}{4}\sqrt{\frac{N}{2M}} - \frac{1}{4} + \mathcal{O}\left(\sqrt{\frac{M}{N}}\right),
\end{equation}
where $M$ is the number of marked nodes. Thus, the algorithm has a quadratic speed-up with respect to a classical search, as the coined quantum walk.

\begin{figure}[hbpt]
	\centering
	\includegraphics[scale=0.58]{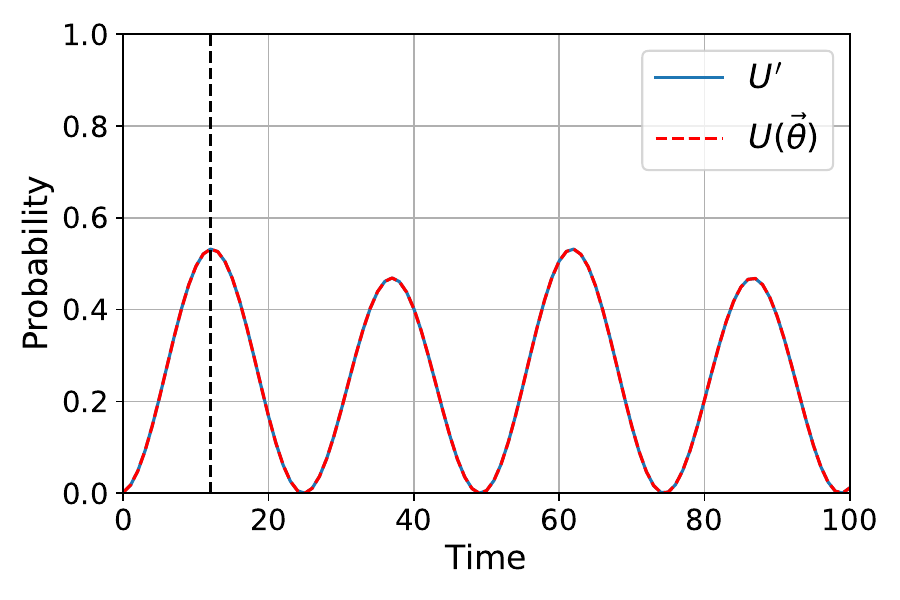}
	\caption{Probability of measuring a marked node versus the time step of a quantum walk search algorithm on a complete graph with $N=1000$ nodes and $M=2$ marked nodes. The nodes have been marked with absorbing vertices in blue, and with a local APR phase $\theta_k=0$ in red. Both algorithms are equivalent, so that the curves overlap. The vertical dashed line indicates the theoretical position of the first maximum, which occurs for $t_{max}=12$.}
	\label{F:search_simulation}
\end{figure}

In order to verify numerically that both models are equivalent, we have simulated the Szegedy's walk with $W_s(\vec{\theta})$ and $W_s'$ for a graph with $N=1000$ nodes and $M=2$ marked nodes. The results are shown in Figure \ref{F:search_simulation}, where we observe that both curves overlap. Moreover, note that the probability oscillates, despite the fact that in the underlying classical walk the walker cannot escape from absorbing vertices. This is due to the unitary character of the evolution, and is related to the ghost links with null transition probability that play an actual role in Szegedy's quantum walk, as discussed in section \ref{sec:Coin}.

An efficient implementation of the update operator $\update$ for the complete graph can be found \cite{Q_circuits}. We can then use the construction in Figure \ref{F:circuit-oracle} to construct a quantum circuit for this quantum walk search algorithm based on an oracle, avoiding the explicit construction of the update operator $\update'$ for the graph with absorbing vertices. Although a form of constructing a circuit for $\update'$ using $\update$ and an oracle was already developed \cite{Sandbichler-master,KMOR15}, this construction needed the application of the oracle two times, so that for the operator $U_s'$ the oracle would be applied four times, instead of two as in our circuit. Furthermore, it needed a controlled version of the operator $\update$, increasing the complexity of the circuit. Therefore, our implementation is simpler. Moreover, our circuit can mark with different phases so that it is not restricted only to absorbing vertices.

\section{Classical simulation with SQUWALS}\label{sec:SQUWALS}

\begin{figure*}
	\centering
	\includegraphics[width=0.75\linewidth]{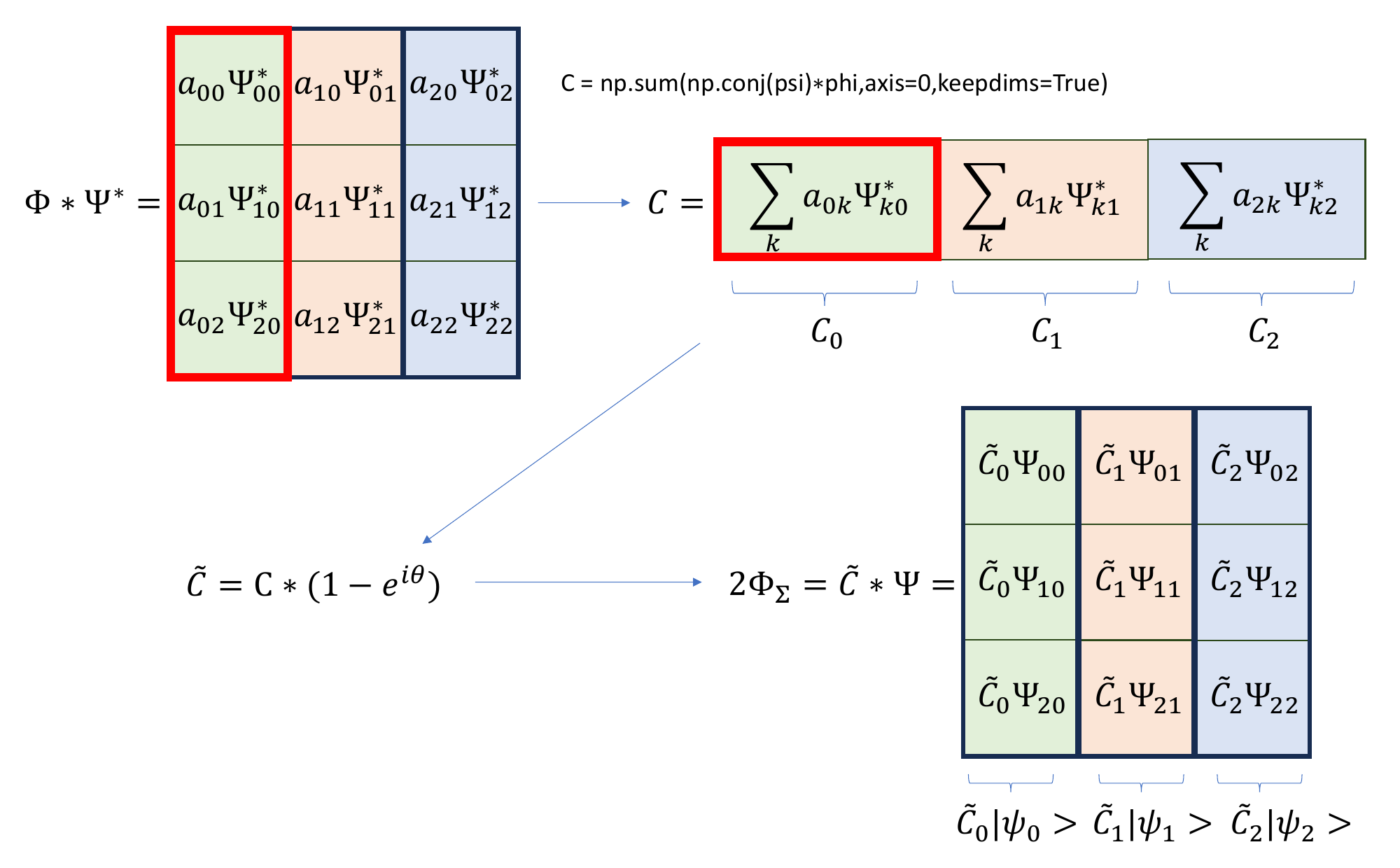}
	\caption{How to obtain the matrix representing the state $2\left|\Phi\right>_\pseudoprojector$ in \eqref{phi_sigma} for a network with $N=3$ nodes. First, we obtain the vector $C$ with the coefficients $C_i$ in \eqref{Ci_1} by the element-wise multiplication between $\Phi$ and $\Psi^*$, and summing the elements of each column. Secondly, the vector $C$ is multiplied element-wise with another vector with the factors $1-e^{\ci\theta_i}$ to obtain the modified coefficients $\widetilde{C}_i$ in \eqref{Ci_2}. Finally, the vector $\widetilde{C}$ is multiplied element-wise with the matrix $\Psi$. The result is a matrix where each column represents the non-null elements of $\widetilde{C}_i\left|\psi_i(\varphi)\right>$, which ends up representing the vector $2\left|\Phi\right>_\pseudoprojector$ as expressed in \eqref{phi_sigma_2}.}
	\label{F:Squwals}
\end{figure*}

The Hilbert space of Szegedy's quantum walk $\mathcal{H}_S$ is of dimension $N^2$. Thus, in order to simulate the walk with a classical computer it is required to construct a $N^2 \times N^2$ matrix representing the unitary evolution operator. This operator has actually $\mathcal{O}(N^3)$ non-null elements for arbitrary transition matrices, so that using a sparse representation of the matrix would provide a classical simulation algorithm scaling as $\mathcal{O}(N^3)$ in both time and memory.

Recently we proposed a novel algorithm that avoids the explicit construction of this matrix, scaling as $\mathcal{O}(N^2)$ for dense transition matrices. Moreover, our algorithm allows the extension with \link phases and a global APR phase. Based on it, we developed a python library called SQUWALS \cite{Squwals}. To simulate the \twnan Szegedy's model we need to modify our algorithm so that it also allows local APR phase extensions, which is not straightforward. In this section we show this modified algorithm, which we have used for all the simulations in this work.

The aim is to find algorithms for simulating the action of the general phase rotation $R(\vec{\theta},\varphi)$ and the swap $S_w$, so that they can be used to simulate any walk operator $U_s(\vec{\theta},\varphi) = S_wR(\vec{\theta},\varphi)$. A generic vector in the Hilbert space $\mathcal{H}_S$ can be written as
\begin{equation}\label{vector}
	\left|\phi\right> = \sum_{i,j=0}^{N-1} a_{ij} \left|i\right>_1 \left|j\right>_2.
\end{equation}
We can represent this $N^2$-dimensional vector as a $N\times N$ matrix $\Phi$, whose elements are
\begin{equation}\label{PHI}
	\Phi_{ij} = a_{ji}.
\end{equation}
Note that using this notation, the column index represents the first register, whereas the row index represents the second register. This is so because if we divide the vector state into blocks corresponding to each state of the computational basis of the first register, then each block corresponds to each column of the matrix state. Using this representation, the action of the swap operator $S_w$ is simulated just transposing the matrix. Nevertheless, the simulation of the phase rotation is quite more complicated.

The phase rotation operator was expressed as $R(\vec{\theta},\varphi) = 2\pseudoprojector(\vec{\theta},\varphi) - \mathbbm{1}$, where $\pseudoprojector(\vec{\theta},\varphi)$ was defined in \eqref{sigma}. Let us denote $\left|\phi\right>_{\pseudoprojector}$ as the result of acting $\pseudoprojector(\vec{\theta},\varphi)$ on the state $\left|\phi\right>$. We need to obtain this vector for the simulation. Indeed, since the factor $2$ in $R(\vec{\theta},\varphi)$ is going to cancel the factor $1/2$ in $\pseudoprojector(\vec{\theta},\varphi)$, we calculate directly the state $2\left|\phi\right>_{\pseudoprojector}$:
\begin{equation}\label{phi_sigma}
2\left|\phi\right>_{\pseudoprojector} = 	2\pseudoprojector(\vec{\theta},\varphi)\left|\phi\right> = \sum_{i=0}^{N-1} (1-e^{\ci\theta_i})C_i \left|\psi_i(\varphi)\right>,
\end{equation}
where the coefficients $C_i$ are obtained as $C_i = \left<\psi_i(\varphi)|\phi\right>$. Let us rewrite the $\left|\psi_i(\varphi)\right>$ states in \eqref{psi_i_extended} as:
\begin{equation}\label{psi_squwals}
	\left|\psi_i(\varphi)\right> = \sum_{k=0}^{N-1}\Psi_{ki}\left|i\right>_1\left|k\right>_2,
\end{equation}
with $\Psi$ being a $N\times N$ matrix whose elements are
\begin{equation}
	\Psi_{ij} = e^{\ci\varphi_{ji}}\sqrt{G_{ij}}.
\end{equation}
This matrix can be obtained taking the element-wise square root of the transition matrix $G$, and multiplying it element-wise with the transpose of the matrix whose elements are the \link phases $e^{\ci\varphi_{ij}}$. Using \eqref{vector} and \eqref{psi_squwals}, the coefficients $C_i$ are obtained as
\begin{equation}\label{Ci_1}
	C_i = \sum_{k=0}^{N-1} a_{ik}\Psi_{ki}^*,
\end{equation}
where the asterisk denotes complex conjugation.

We can calculate the coefficients $C_i$ in a vectorized way avoiding a \textit{for} loop. We multiply element-wise the matrix state $\Phi$ with the conjugate of the matrix $\Psi$, and add across the rows of the resulting matrix, obtaining a vector $C$ with the coefficients. In the upper panel of Figure \ref{F:Squwals} we show an example for a network with $N=3$ nodes.

We define the modified coefficients $\widetilde{C}_i$ as:
\begin{equation}\label{Ci_2}
	\widetilde{C}_i = (1-e^{\ci\theta_i})C_i,
\end{equation}
so that equation \eqref{phi_sigma} can be rewritten as
\begin{equation}\label{phi_sigma_2}
	2\left|\phi\right>_{\pseudoprojector} = \sum_{i=0}^{N-1} \widetilde{C}_i \left|\psi_i(\varphi)\right>.
\end{equation}
The vector $\widetilde{C}$ representing the new coefficients is obtained creating a vector with the factors $1-e^{\ci\theta_i}$ and multiplying it element-wise with the vector $C$. We can use the broadcasting feature of NumPy \cite{NumPy} to multiply element-wise the row vector $\widetilde{C}$ with the matrix $\Psi$. The resulting matrix has in the $i$-th column the non-null elements of $\left|\psi_i(\varphi)\right>$ multiplied by the coefficient $\widetilde{C}_i$. Thus, it results in the matrix representing the vector $2\left|\phi\right>_\pseudoprojector$. The whole procedure is shown in Figure \ref{F:Squwals} for the example with $N=3$ nodes.

Now we have the action of the operator $2\pseudoprojector(\vec{\theta},\varphi)$, the action of $R(\vec{\theta},\varphi)$ is trivially obtained by an element-wise matrix subtraction of the initial state, because using \eqref{general_R} we have
\begin{equation}\label{subtraction}
	R(\vec{\theta},\varphi)\left|\phi\right> = 2\left|\phi\right>_\pseudoprojector - \left|\phi\right>.
\end{equation}

In all the procedures needed to simulate the Szegedy's evolution operator the bigger mathematical objects that intervene are $N \times N$ matrices. Thus, the memory complexity of the algorithm scales as $\mathcal{O}(N^2)$. Moreover, since the multiplications are performed element-wise, they involve at most $N^2$ operations. Thus, the time complexity is also expected to scale as $\mathcal{O}(N^2)$.

\section{Conclusions}\label{Conclusions}

We have reviewed the formulation of Szegedy's quantum walk, and the current phase extensions in the literature based on \link phases and a global arbitrary phase rotation (APR), to generalize it including local APR phases. Therefore, we have defined different Szegedy's model depending on each kind of phase extensions, and the \twnan Szegedy's quantum walk including all the phase extensions.

Based on the quantum circuit construction of the standard Szegedy's walk, we have shown how a circuit can be modified in order to include phase extensions. In general cases where all the phase values would be different, the circuits would be inefficient. However, the same as happens with the transition matrix for the standard model, the phases can be distributed in some symmetric patterns so that an efficient implementation is possible. For example, if there is a small set of nodes with a different local APR phase, or there are only two local APR phases values distributed between even and odd nodes.

It is known that under some circumstances the coined quantum walk model and the standard Szegedy's quantum walk are equivalent. We have reviewed this equivalence considering the different phase extensions and found that they can be used to cast a wider set of coins into Szegedy's model. Therefore, we have formulated some lemmas about the conditions that need to satisfy the coin operators in order to establish an equivalence with the different Szegedy's models. Of course, the \twnan Szegedy's model can host the bigger set of coins.

We have shown how the two quantum walk models can be cast into each other using an example on the 1D line. On the one hand, in the case of the standard Szegedy's walk based on the undirected graph, the equivalent coin is the Pauli $X$ matrix. On the other hand, a Hadamard coin walk is cast into a Szegedy's walk on an unbiased weighted graph, where the probability is bigger for jumping to the right. Simulations show that both coins are not equivalent, despite the fact that the Hadamard coin was though to be a sensible quantization of the classical walk on the undirected graph. Furthermore, we have seen an example complex coin that cannot be cast into standard Szegedy's model, but can be cast into the \twnan model using the phase extensions. This coin produces the same results as the Hadamard coin when the coins are the same for all nodes. Nevertheless, when a coined walk is quantized using both coins simultaneously, distributing them between even and odd nodes, a new different walk is obtained. This result is quite surprising, and shows that coins that \apriori seems to be equivalent, are not under different circumstances. Furthermore, based on the symmetric pattern of distribution of the extended phases, we have managed to construct a quantum circuit for this coined walk with two coins.

An important application of the local APR phases is that they can be used to mark nodes in a graph, so that the quantum evolution treats them in a different manner. We have seen how a quantum circuit based on oracles can be constructed, so that there is no need to modify the structure of the graph and it is more efficient that previously known constructions. Quantum search algorithms are examples where nodes are marked. We have shown that marking with a null local APR phase is totally equivalent to marking with absorbing vertices in the case of using the double Szegedy operator $W_s$. When using the single operator $U_s$ the same results are obtained after measuring in the computational basis, although both operators are not exactly equal.

Although the $-\mathbbm{1}$ coin cannot be cast into the standard Szegedy's walk with the single operator $U_s$, we have seen that we can cast it for the double operator $W_s$. This result is intriguing, as shows that when considering the double operator the set of coins that can host is wider, not needing any complex phase extension. More research about this fact is needed in the future.

Finally, we have provided a classical simulation algorithm for the \twnan Szegedy's quantum walk, including local APR phases in our previous algorithm implemented in SQUWALS \cite{Squwals}. This algorithm scales as $\mathcal{O}(N^2)$ in both time and memory requirements, so that is more efficient than previous algorithms scaling as $\mathcal{O}(N^3)$.

A quantum PageRank algorithm using Szegedy's walk with global APR was the first algorithm showing the utility of the phase-rotation extension \cite{APR}. The \link phases have also been considered for problems of state transfer in quantum walks \cite{State_transfer}. In the future, it would be interesting to study further applications considering all the phase extensions at the same time. For example, taking into account that local APR phases act at the level of nodes, and \link phases at the level of edges, there could be a relation with gauge symmetries. Moreover, we could include the phase extensions considering the twisted swap operator \cite{Twisted}, and study how our \twnan walk can be mapped to the general quantum singular value transform (QSVT) framework \cite{QSVT_1,QSVT_2}, which is an interesting but open problem.

\section{Data Availability Statement}\label{Data}

The library SQUWALS with the new simulation algorithm and a tutorial are available on GitHub: \url{https://github.com/OrtegaSA/SQUWALS-repo}.

\section*{Acknowledgments}

We acknowledge the support from the Spanish MINECO grants MINECO/FEDER Projects,  PID2021-122547NB-I00 FIS2021, the “MADQuantum-CM" project funded by Comunidad de Madrid and by the Recovery, Transformation, and Resilience Plan – Funded by the European Union - NextGenerationEU and Ministry of Economic Affairs Quantum ENIA project. This work has also been financially supported by the Ministry for Digital Transformation and of Civil Service of the Spanish Government through the QUANTUM ENIA project call – Quantum Spain project, and by the European Union through the Recovery, Transformation and Resilience Plan – NextGenerationEU within the framework of the Digital Spain 2026 Agenda. M. A. M.-D. has been partially supported by the U.S.Army Research Office through Grant No. W911NF-14-1-0103. S.A.O. acknowledges support from Universidad Complutense de Madrid - Banco Santander through Grant No. CT58/21-CT59/21.

\bibliography{MiBiblio}
\bibliographystyle{unsrt}

\end{document}